\documentclass[]{aa}
\usepackage{graphicx}
\usepackage{txfonts}
\usepackage{xcolor}
\usepackage{ulem}
\usepackage{subfigure}
\bibpunct{(}{)}{;}{a}{}{,}

\newcommand{\SC}[1]{\textcolor{black}{#1}} 
\newcommand{\SCC}[1]{\textcolor{black}{#1}} 
\newcommand{\MVSC}[1]{\textcolor{black}{#1}} 
\newcommand{\qu}[1]{\textcolor{black}{#1}} 
\newcommand{\COI}[1]{\textcolor{black}{#1}} 
\newcommand{\LEt}[1]{\textcolor{black}{#1}} 

\newcommand{\al}[1]{\textcolor{black}{#1}} 

\newcommand{\Msun}{M$_{\odot}$}

\newcommand{\Mstar}{M$_{\star}$}

\newcommand{\kms}{km~s$^{-1}$}
\newcommand{\RDUST}{R$_{ \rm dust}$}
\newcommand{\RCO}{R$_{ \rm CO}$}

\begin{document} 

%

   \title{ALMA reveals a large structured disk and nested rotating outflows in DG~Tau~B}

   \subtitle{}

   \author{A. de Valon
          \inst{\ref{inst1}}
          \and
          C. Dougados 
          \inst{\ref{inst1}}
          \and
          S. Cabrit
          \inst{\ref{inst2}\and\ref{inst1}}
          \and
          F. Louvet
          \inst{\ref{inst3}}
          \and 
          L. A. Zapata
          \inst{\ref{inst4}}
          \and 
          D. Mardones
          \inst{\ref{inst3}}
          }

   \institute{Univ. Grenoble Alpes, CNRS, IPAG, 38000 Grenoble, France \label{inst1}
             \and
             PSL University, Sorbonne Universit\'{e}, Observatoire de Paris, LERMA, CNRS UMR 8112, 75014 Paris, France \label{inst2}
             \and 
             Departamento de Astronomía de Chile, Universidad de Chile, Santiago, Chile
             \label{inst3}
             \and
             Instituto de Radioastronomía y Astrofísica, Universidad Nacional Aut\'onoma de M\'exico, P.O. Box 3-72, 58090, Morelia, Michoac\'an, M\'exico
             \label{inst4}
            }

   \date{}

 
  \abstract
   {We present Atacama Large Millimeter Array (ALMA) Band 6 observations at 14-20~au spatial resolution of the disk and CO(2-1) outflow around the Class I protostar DG~Tau~B in Taurus. The disk is very large, both in dust continuum (R$_{\rm eff,95\%}$=174~au) and CO (R$_{CO}$=700~au). It shows Keplerian rotation around a 1.1$\pm$0.2~\Msun~central star \SC{and two dust  \SCC{emission} \al{bumps} at $r$ = 62 and 135~au. \SCC{These results confirm that large structured disks can form at an early stage where residual infall is still ongoing}. The redshifted CO outflow \SCC{at high velocity} shows a striking hollow cone morphology  \SC{out to 3000 au} with a \COI{shear}-like velocity structure within the cone walls. \SCC{These walls coincide with the scattered light cavity, and they appear to be rooted within $<$~60~au in the disk}.
   We confirm \COI{their} global average rotation in the same sense as the disk, with a specific angular momentum  $\simeq$ 65~au~\kms. The mass-flux rate of \al{1.7-2.9}~$\times$~10$^{-7}$\Msun~yr$^{-1}$ is \SC{\al{35$\pm$10} times that in the atomic jet}. 
   We also detect a wider and slower outflow component surrounding this inner conical flow, which also  rotates in the same direction as the disk. Our ALMA observations therefore demonstrate that the inner cone \SCC{walls}, \SCC{and the associated scattered light cavity,} do not trace the interface with infalling material, which is shown to be confined to much wider angles ($> 70^{\circ}$). 
   \qu{The properties of the conical walls are suggestive of the interaction between an episodic inner jet \LEt{or} wind with an outer disk wind, or of a massive disk wind originating from 2-5 au}. However,
   \SCC{ further modeling is required to establish their origin.}
   In either case, such massive outflow may significantly affect the disk structure and evolution}.}

   
   
   

    \titlerunning{ALMA observations of DG Tau B disk and CO outflow }
   \maketitle
%

\section{Introduction}

Jets and outflows are ubiquitous at all stages of star formation. However, their exact origin and  impact on mass and angular momentum extraction are still fundamental open questions \citep{frank_jets_2014}. 
Slow molecular outflows are traditionally interpreted as swept-up material, tracing the interaction of the jet or a wide-angle wind with the infalling envelope and parent core \citep[eg. in ][]{zhang_alma_2016}. 
Recent interferometric observations
with the Atacama Large Millimeter Array (ALMA) have
revealed, in a few cases, rotation levels suggesting that material directly ejected from the disk, through photo-evaporation or magneto-centrifugal processes,  might also perhaps contribute to the formation of  low velocity molecular cavities on small scales.
These rotating cavities have been observed at all stages from Class 0 \citep[e.g., HH212, IRAS 4C in][]{tabone_interaction_2018,zhang_rotation_2018}, Class I \citep[e.g., TMCIA in ][]{bjerkeli_resolved_2016}, and Class II stars \citep[HH30 in ][]{louvet_hh30_2018}.
If present, such disk winds could play a crucial role in the transport of angular momentum through the disk and the final dispersion of gas.
Studies at high-angular and spectral resolution of objects with weak or no infalling envelopes with ALMA
are crucial to help discriminate between an envelope and a disk origin for the rotating outflow material. 

DG~Tau~B is a highly reddened Class I/II protostar located in the dark globule B217 in the Taurus complex \citep{schmalzl_star_2010}. 
GAIA DR2 parallaxes of optically visible young stars in the same globule\footnote{excluding DG Tau~A where bright nebulosity introduces an unusually high  measurement error for its G magnitude \citep{brown_gaia_2018}} 
suggest a distance $\simeq$140~pc, which we adopt thereafter. 
This object drives an asymmetric bipolar jet at PA=122$^{\circ}$ \citep{mundt_collimation_1991}, which was  first discovered by \citet{mundt_jets_1983} and studied in detail by \citet{podio_tracing_2011}. 
%
Asymetric conical scattering nebulae have also been observed at optical and near-infrared wavelengths \citep{padgett_hst/nicmos_1999} with \SC{half} opening angles of respectively 
$42^\circ$ (blue lobe) and $22^\circ$ (red lobe). 
A slow asymmetrical molecular outflow \SCC{was} first mapped in CO by \citet{mitchell_dg_1997}, \SCC{which was mainly} associated to the receding atomic jet. \citet{mottram_outflows_2017} show that both CO lobes extend up to at least 60$^{\prime\prime}$ (8400 au) with opening angles similar to the infrared scattering nebulae. They also show the absence of HCO+ emission on source, which is typical for young stars with weak envelopes \citep{van_kempen_nature_2009}. \citet{zapata_kinematics_2015} studied DG~Tau~B with the Submillimeter Array (SMA)  at 2$^{\prime\prime}$ angular resolution and they report velocity asymmetries across the receding conical CO outflow interpreted as rotation signatures. \al{They propose that the massive redshifted conical outflow traces entrained material from the rotating parent core.}

In this paper we present continuum, $^{12}$CO(2-1), $^{13}$CO(2-1), and C$^{18}$O(2-1) observations acquired with ALMA of the DG~tau~B disk and outflow, which increase the angular resolution by more than a factor of 10 with respect to \COI{the} previous study \COI{of \citet{zapata_kinematics_2015}}.
In Sect. \ref{Obs} we describe the observations and data reduction. In Sect. \ref{Results} we present 
new results 
revealing a large and structured disk as well as a rotating outer wide angle component, surrounding the inner conical flow. In Sect. \ref{Discussion} we discuss 
\SC{new implications for the origin of disk structure and small-scale conical CO outflows.} 

\section{Observations and data reduction}
\label{Obs}

\SC{DG Tau B was observed \al{in band 6} during ALMA Cycles 3 and 5 with \al{three 12m array configurations covering} a baseline range of 11--6500k$\lambda$ (15--8500m). \qu{The details of the observations are presented in Table \ref{t1}.} }
The spectral set-up \al{was similar for each observation and} included three spectral windows centered on the $^{12}$CO(2-1), $^{13}$CO(2-1), and C$^{18}$O(2-1) \COI{emission} lines, each with a \al{native} spectral resolution of 122~kHz, which we rebinned to 244~kHz ($\approx 0.3$~\kms) to improve sensitivity. An additional spectral window with 2~GHz bandwidth centered at $\simeq$ 232~GHz 
was used to map the continuum. 
\al{The three configurations were entirely reprocessed and combined both in continuum and $^{12}$CO to improve the signal-to-noise ratio (S/N) and the quality of the data} as well as to \SCC{ greatly reduce the residual sidelobes in the final cleaned images}.
Data \al{calibration was performed} using the Common Astronomy Software Application \citep[CASA, see][]{mcmullin_casa_2007} version 4.5.3 for Cycle 3 observations and 5.1.1 for the Cycle 5 observations.   \al{Flux calibrators were manually checked and the flux calibration uncertainty was estimated to be 20\%}.
For each individual exposure, phase self-calibration was \al{computed on the high signal-to-noise} \al{continuum and applied to both the} continuum and $^{12}$CO data. \al{The Cycle 5 continuum} \SCC{peak} (\al{centered at} $\alpha_{J2000}$: 04$^h$27$^m$02$^s$.573, $\delta_{J2000}$: +26$^{\circ}$05$^{\prime}$30$^{\prime\prime}$.170) \al{was defined} as a phase reference. 
Relative phase shifts between all different exposures from the three configurations were automatically corrected by the self-calibration
procedure.
\al{Final calibrated \LEt{data sets} from the three different configurations were combined with their default weights}. Before combination, we checked that the average weights of each \LEt{data set} were similar.
Joint deconvolution was achieved by the CASA tclean procedure
using \al{the Hogbom algorithm and} Briggs weighting scheme with a robust parameter of -1 for the continuum data and 0.5 for the CO. The resulting continuum emission map has a synthesized beam of 0.07$^{\prime\prime}$x0.11$^{\prime\prime}$ with a root mean square (rms) noise
of $50~\mu$Jy/beam. The final $^{12}$CO(2-1) datacube has a synthesized beam of 0.13$^{\prime\prime}$x 0.18$^{\prime\prime}$ 
with an rms noise of 2 mJy/beam per 0.3~\kms spectral channel. For the more compact $^{13}$CO(2-1) 
and C$^{18}$O(2-1) emission, we used the \al{pipeline reduced datacubes from the compact C36-2/3 and intermediate} C40-5 configurations, with resulting beams of \al{$0.97^{\prime\prime} \times 0.78^{\prime\prime}$} 
and $0.23^{\prime\prime} \times 0.34^{\prime\prime}$, respectively, and rms noise per 0.3~\kms~spectral channel of 6 and \qu{\al{5} mJy/beam, respectively}. \al{Our largest recoverable scale is 11$^{\prime\prime}$}.
\section{Results \al{and analysis}}
\label{Results}


\subsection{A large and structured disk}

\begin{table*}
\caption{\label{t2} Best fit parameters to the disk continuum emission profile}
\centering
\resizebox{\textwidth}{!}{\begin{tabular}{cccc|ccc|ccc}
\hline
\hline
\multicolumn{4}{c|}{Main profile} & \multicolumn{3}{c|}{Inner ring} & \multicolumn{3}{c}{Outer ring} \\
A     &  $r_t$    &  $\gamma$ &   $\beta$ &  $ B_1 $  &  $R_1 $    &    $\sigma_1 $  &  $ B_2$ &  $R_2 $    &   $ \sigma_2$   \\
(mJy/beam)& (AU) & & &(mJy/beam) & (AU) & (AU)& (mJy/beam) & (AU) & (AU) \\
\hline
$30\pm1$ & $18\pm 1$ & $0.07 \pm 0.02$ &$0.7 \pm 0.2$ & $1.5 \pm 0.1$ & $62.3\pm 0.5$ & $4.7 \pm 0.5$ & $0.4 \pm 0.02$ & $135 \pm 1$ & $22 \pm 1$ \\
\hline
\end{tabular}}
\end{table*}

\begin{figure*}
    \resizebox{\hsize}{!}{\includegraphics{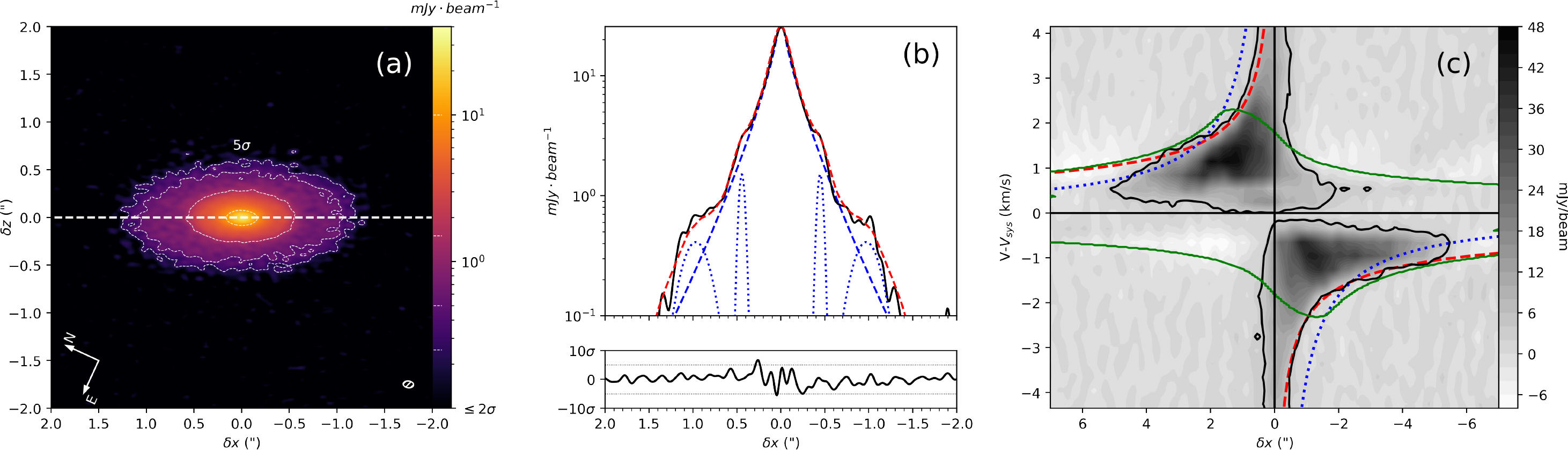}}
    \caption{(a) Continuum map at 232~GHz in logarithmic scale. White dotted contours show the 5, 10, 30, and 200 $\sigma$ levels ($\sigma$=50 ~$\mu$Jy/beam). The white dashed line shows the disk \al{major axis at PA=25.7$^{\circ}$}. (b) Continuum intensity profiles: \al{Cut} along PA=25.7$^{\circ}$ (in black) \al{and deprojected and azimuthally averaged (in red)}. The tapered power law central component and the two Gaussian components, derived from the fitting and before beam convolution, are shown in \al{blue} dashed and dotted lines, respectively. Bottom panel shows the fit residuals and $\pm 5\sigma$ limit (dotted lines). (c) Position-velocity diagram in $^{13}$CO along PA= 25.7$^{\circ}$~in a pseudo-slit of \al{4}$^{\prime\prime}$ width. The $3\sigma$ contour is shown in black. \al{The curves show maximum expected line-of-sight velocities $V_{\rm max}$ for three different models: a thin disk in Keplerian rotation \SCC{around} $M_\star=1.1 M_{\odot}$ at an inclination of $i=63 ^{\circ}$ (red dashed line), \SCC{pure rotation with constant angular momentum} $V_{\rm max} \propto 1/r$ (blue dotted line), and a rotating and infalling \SCC{bipolar} shell (see Sect. \ref{OUTER_OUTFLOW}) with Rd=300~au, $\theta_0=70^{\circ}$ \COI{at an inclination of $i=63 ^{\circ}$} (green solid line).}
    }
    \label{fig:Disk}
\end{figure*}

Figure~$\ref{fig:Disk}$a shows the 230~GHz continuum image. The continuum emission 
is elliptical and \al{centrally peaked}, indicating a dominant contribution from the circumstellar disk. The peak intensity is 26~mJy/beam (80 K)
and the integrated intensity is $430\pm 30$~mJy. The difference with Zapata (590 +/- 30 mJy) and Guilloteau (531 mJy) is within the calibration uncertainties of 20\%. We fit an elliptical Gaussian with the CASA function IMFIT to derive a position axis (PA) of $25.7 \pm 0.3 ^\circ$ for the disk major axis, which is consistent with the previous determinations of 24~$\pm$~1$^\circ$ by \citet{guilloteau_dual-frequency_2011} and 30~$\pm$~5$^\circ$  by \citet{zapata_kinematics_2015}. 
Assuming that the disk is axisymetric and that the millimetric continuum emission is mainly located in the midplane, we also derived the disk axis inclination to the line of sight at $i$= 63\degr $\pm 2^\circ$, which is consistent with the estimate of 64\degr $\pm 2^\circ$ by \citet{guilloteau_dual-frequency_2011}.

Figure $\ref{fig:Disk}$b shows the intensity profile of the continuum along the disk major axis. 
The effective radius encompassing 95\% of the \COI{deprojected} continuum flux \COI{distribution} $\int I_\nu(r) 2\pi r dr$ is $R_{\rm eff,95\%}$=1.24$^{\prime\prime}$=174~au.
The profile shows a centrally peaked component with two symmetrical emission bumps.
\MVSC{The deprojected and azimuthally averaged radial profile (dashed red line
in Fig. $\ref{fig:Disk}$b) shows that this feature is axisymmetric, albeit with a lower contrast due to the large inclination of the system.} 
\al{ We follow the modeling approach developed by \citet{long_gaps_2018} for similar intensity profiles in \SCC{Class II} disks}. 
We fit the intensity profile with an exponentially tapered power-law for the central component and 
two Gaussian components for the \al{emission bumps} (see Appendix~\ref{sec:BestFit} for details). 
The best fit \al{components} are shown in Fig. \ref{fig:Disk}b together with
\SC{the fit residuals;} the parameters are given in Table~1. \al{The derived model radial intensity profile (before beam convolution) is shown in Fig.~\ref{fig:profil_sconv}.} 
\al{The possible origins of the observed structures are discussed in Sect. \ref{sec:structure_disk}.}

Figures~\al{\ref{fig:13CO_channels}} \& \ref{fig:C18O_channels} present channel maps in \al{$^{13}$CO} and C$^{18}$O. They exhibit the characteristic "double loop" pattern and trace the upper and lower surfaces of  disks \al{in Keplerian rotation}, as in IM~Lupi \citep{pinte_direct_2018}, for example. The C$^{18}$O emitting layers appear to be at $z/r \simeq$ 0.3, which is similar to IM~Lupi. \al{A contribution from the redshifted outflow is also observed in $^{13}$CO at $\delta z < 5"$ and for ($V-V_{\rm sys}$)~$\leq$~2.5~\kms~(see Fig. \ref{fig:13CO_channels})}. 
Figure~\ref{fig:Disk}c shows the $^{13}$CO position-velocity (PV) diagram perpendicular to the flow axis, averaged over $-2^{\prime\prime} < \delta z < 2^{\prime\prime}$ to increase the S/N. \al{We also show maximal projected velocities expected for a disk in Keplerian rotation as well as a rotating infalling shell model (see 3.2.2 for more details on this model)}. 
The kinematics are clearly \al{best fit by Keplerian} rotation \al{out to r=5$^{\prime\prime}$}. However, faint redshifted emission is also seen at $-2'' < \delta x < 0$ (the wrong direction for rotation), requiring residual infall on the front side of the disk \SCC{(see green curve in Fig.~\ref{fig:Disk}c)}. \al{The fact that we do not see a symmetric}
\SCC{infall signature} \al{at negative velocities and $\delta x > 0$ may come from asymetrical infall as seen, for example, in L1489-IRS by \citet{yen_alma_2014}}.
We thus concentrate on the blue wing of the PV diagram, which is less affected by \al{outflow} contributions. It shows a clear Keplerian profile out to $r=5^{\prime\prime}$, corresponding to a disk radius of $R_{\rm CO}$=700~au. The same disk size is indicated in $^{12}$CO (Fig. $\ref{fig:Moneyplot}$).
We estimated the stellar mass by using the model of a thin disk in Keplerian rotation. 
With an inclination $i$ = 63\degr, the best fit to the 3$\sigma$ contour of the $^{13}$CO PV diagram is found for a systemic \LEt{local standard of rest (}LSR\LEt{)} velocity $V_{\rm sys}=6.3\al{5} \pm \al{0.05}$ \kms~and a stellar mass \Mstar~=~1.1~$\pm$~0.2~\Msun. 
This is the most accurate determination so far of the \al{proto}stellar mass for DG~Tau~B. 

\subsection{$^{12}$CO redshifted lobe: Nested rotating outflows}

\begin{figure}
    \resizebox{\hsize}{!}{\includegraphics{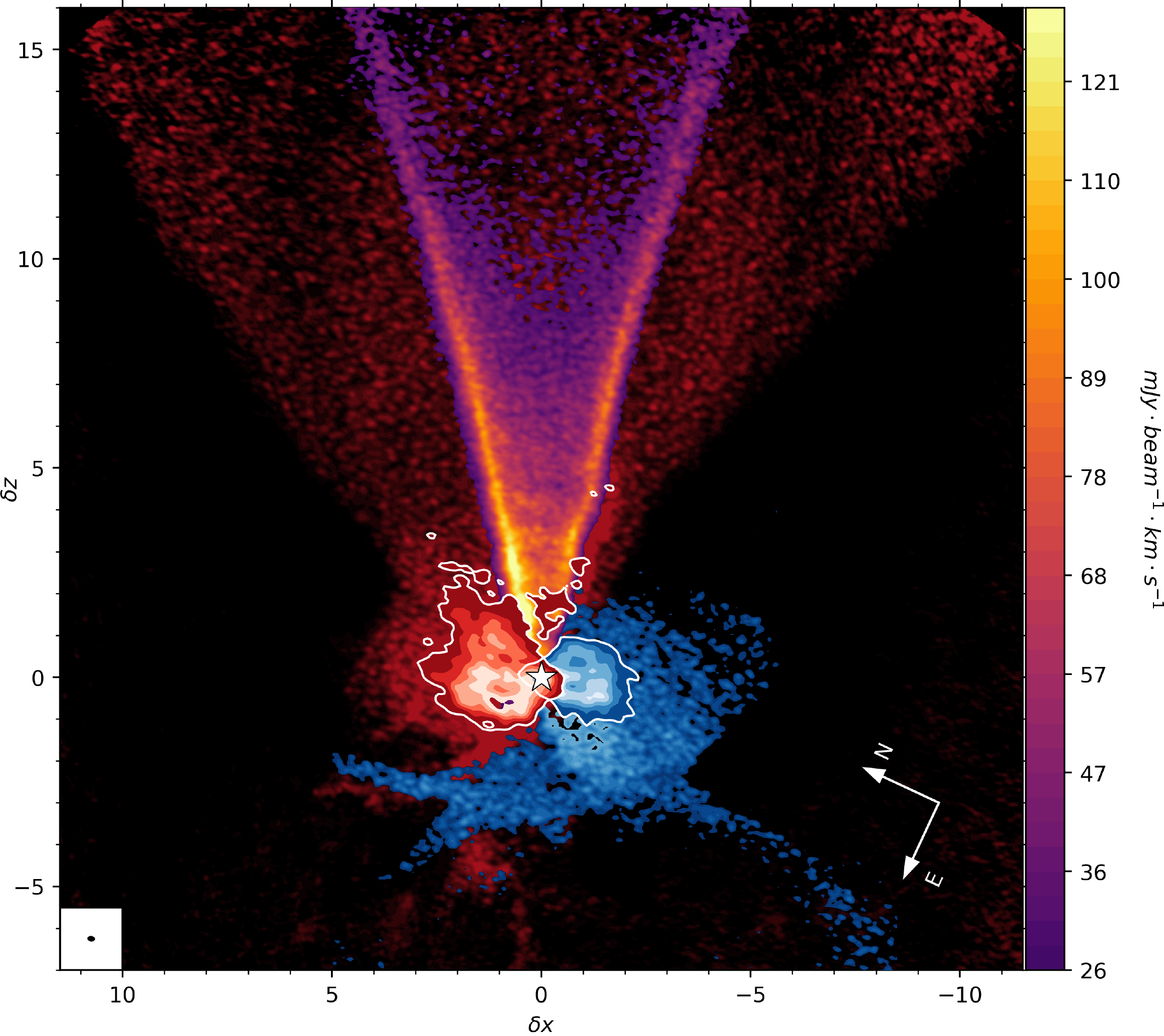}}
    \caption{Summary of the various $^{12}$CO(2-1) kinematical components in DG~Tau~B: The inner conical flow (integrated from $(V-V_{\rm sys})$~=~+2.15 to +8~\kms) is shown in purple to yellow shades (floor at \al{2\COI{6}~\COI{m}Jy/beam}~\kms). \al{The individual low-velocity $^{12}$CO(2-1) channel maps at $(V-V_{\rm sys})=1.19$~\kms~ (resp. -1.03~\kms~)} are shown in red (resp. blue) shades (floor at 3$\sigma$ with $\sigma$~=~2~mJy/beam). Inside the white contours, the $^{13}$CO(2-1) channel maps are shown at $(V-V_{\rm sys}) =  1.45$~\kms~\al{in red and $(V-V_{\rm sys}) = -1.55$~\kms~in blue,} highlighting the upper and lower disk surfaces (floor at 3$\sigma$ with $\sigma$~=~5~mJy/beam).}
    \label{fig:Moneyplot}
\end{figure}

\al{Figure~\ref{fig:Moneyplot} illustrates the change in morphology of the CO redshifted lobe emission as a function of velocity. \COI{Individual channel maps are shown in Fig. \ref{fig:CO_channels}}. At large velocities ($V-V_{\rm sys} = 2-8$ \kms), channel maps show a striking limb-brightened conical morphology with an almost constant opening angle out to at least z=10$^{\prime\prime}$=1400~au.}
\al{At lower velocities ($V-V_{\rm sys} = 1.2-1.8$ \kms), a  wider, \SCC{thicker,} nonconical component dominates the emission. The transition \SCC{in morphology} 
between low and mid-high velocities can also be observed on the transverse PV diagrams in Fig.~\ref{fig:XVcone} where the slope of the \SCC{velocity gradient steepens} at around $V-V_{\rm sys} \simeq 2-2.5$~\kms. Although it is not clear whether these two kinematical components trace two physically distinct mechanisms, we discuss their specific properties separately below.}

\subsubsection{Inner rotating conical flow \SCC{at $(V-V_{\rm sys}) > 2$~\kms}}

\begin{figure*}
    \centering
    \resizebox{\hsize}{!}{\includegraphics{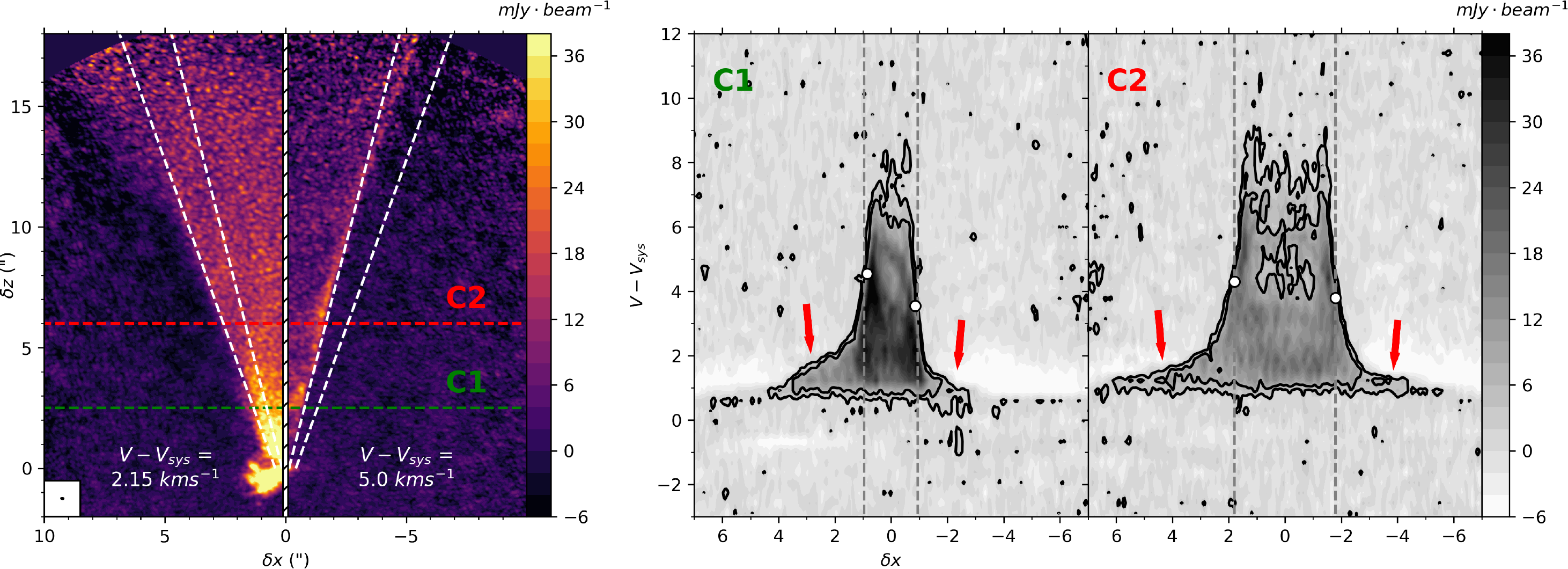}}
    \caption{\SCC{Left panel:} $^{12}$CO channel maps at $(V-V_{\rm sys})=$ \SCC{2.15 }\kms~and $(V-V_{\rm sys})=\qu{5.00}$~\kms. The two white dashed lines trace the outer and inner cones (see text). \COI{Only one side of the outflow is shown in each of the two channel maps}. The color scale is the same for the two channel maps. Right panel: \al{Transverse} PV diagrams at $\delta z =2\al{.5}"$ and $\delta z =6"$ obtained along the C1 and C2 cuts shown on the channel maps and \al{averaged on a slice of $\Delta z = 0.2"$. The black contours trace the 2$\sigma$ and 5$\sigma$} levels.
    The two \al{white} points represent the \al{symmetric edges} of the PV diagrams at $\delta x = \pm 0.\al{8}5"$ and $\delta x = \pm 1.8"$, respectively. Their \al{velocity} difference gives an estimate of the rotation velocity (see text). 
    }
    \label{fig:XVcone}
\end{figure*}

\al{The channel maps presented in Fig.~\qu{\ref{fig:XVcone}} illustrate the conical morphology of the emission at $\delta z \leq 10"$ and 
\COI{at} $(V-V_{\rm sys})\qu{= 2}$~\COI{and}~$\qu{5}$}~\kms~\al{and they show a decreasing opening angle with an increasing line-of-sight velocity}. 
\MVSC{The outer cone is defined from the channel map at 2.15~\kms, which corresponds to the \SCC{lowest} velocity channel where the conical component dominates the emission until $\delta z \approx 10"$.}  \al{The derived outer semi-opening angle of $\simeq$18$^{\circ}$ strikingly coincides with the opening of the }near-infrared (NIR) scattered light cavity
\SCC{observed} by \citet{padgett_hst/nicmos_1999}. 
\SCC{Figure~\al{\ref{fig:XVcone}} shows that}
the cone opening angle decreases 
to 12$^\circ$ at $(V-V_{\rm sys}) = \qu{5}.0$~\kms,
\al{and Fig. \ref{fig:CO_channels} shows that it stays constant at higher velocities}.
This \al{strong velocity stratification} \SCC{and hollow cone geometry} is confirmed by the shape of the transverse PV diagrams, which were constructed by positioning a pseudo-slit perpendicular to the flow axis (\SCC{right panels of} Fig.~ \ref{fig:XVcone}). \al{The outer contours of the PV diagrams decrease in radii with increasing projected velocities, and then they become almost vertical at the highest velocities \SCC{$(V-V_{\rm sys})$~$\geq$ 5~\kms}, defining a limiting inner cone.}
Because the velocity gradient in the conical component is limited to a rather thin shell and narrow range of angles, this suggests a \SCC{"shear-like"} velocity structure within the cone walls with faster material \SCC{on the internal side}. 

By extrapolating the conical morphology \SCC{fit \al{by \LEt{the naked }eye} to the channel maps of Fig.~\ref{fig:XVcone}} \al{at $\delta z < 10"$}, we derived upper limits to the anchoring radii in the range of 10 to 60~au for the inner and outer cone, respectively. No important dynamical or structural changes are observed along the outflow cone over its extent, \qu{except for a small increase in the opening angle at $\delta z$ > $10"$}, suggesting a quasi steady-state flow.

The PV cuts also reveal a systematic velocity difference between symmetric positions at $\pm r$ from the flow axis, especially at low heights above the disk (see white circles in Fig.~\ref{fig:XVcone}). Assuming axi-symmetry, this difference can only be due to rotation. We thus confirm the finding of \citet{zapata_kinematics_2015} that the conical \SCC{redshifted outflow} component rotates coherently with altitude and in the same direction as the disk. We observe a velocity shift $\Delta$V=~1.\al{0}~\kms~ at z = 2\al{.5}$^{\prime\prime}$ at a radius of r=0.\al{8}5$^{\prime\prime}$(=1\al{20}~au), indicating a typical specific angular momentum of $\simeq$ 65~au~\kms~\al{for $V_Z$ $\approx$ 9 \kms.} A similar value is found at other positions along the cone (Fig.~\ref{fig:XVcone}). 

\SCC{Assuming a steady flow,} \al{the mass flux along the conical structure is estimated from  $(V-V_{\rm sys}) = 2.15$ to 8~\kms~at $\dot{M} = 1.7-2.9\times 10^{-7}~M_{\odot}~yr^{-1}$ depending on the opacity hypothesis (see Appendix \ref{sec:Mass_flux} for more details).}
The derived mass flux is \SCC{$35\pm 10$} times larger than the mass flux of the receding atomic jet of $6.4 \times 10^{-9}$~M$_{\odot}$~yr$^{-1}$ estimated by \citet{podio_tracing_2011}, making it of the same order as the mass accretion rate onto the star if a standard ejection/accretion ratio of \al{0.01}--0.1  \al{\citep{hartigan_mass-loss_1994}} is adopted for the atomic jet. 
On the other hand, the linear momentum flux for the inner CO conical outflow is on the same order of magnitude as that of the jet \al{(\al{two to four} times higher depending on the \qu{opacity hypothesis})}. \SC{Similar results were found in the smaller conical outflow of HH30 \citep{louvet_hh30_2018}} \SCC{and in the inner molecular outflow shells of the Class 0/I source HH46-47 \citep{zhang_episodic_2019}}

\subsubsection{Outer low-velocity wide-angle outflow}
\label{OUTER_OUTFLOW}
\begin{figure*}
    \centering
    \resizebox{\hsize}{!}{\includegraphics{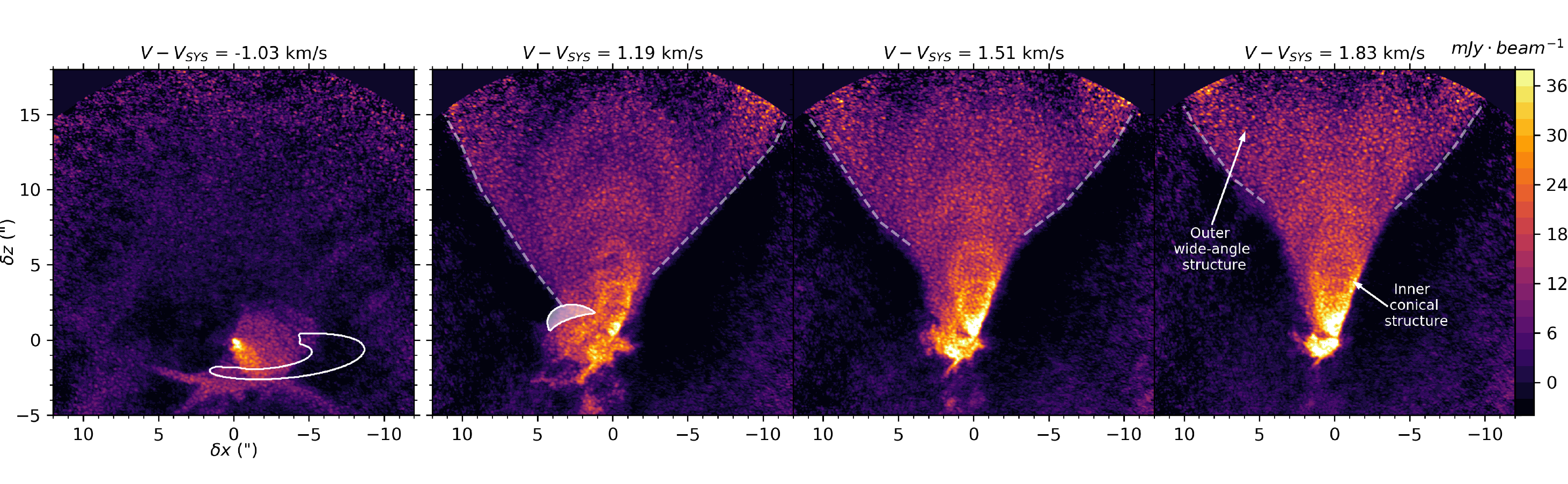}}
    \caption{$^{12}$CO \al{individual} channel maps at different line-of-sight velocities. The gray filled white contours trace the model of an infalling shell with $r_d = 700 $~au and $\theta_0=70^{\circ}$. The contours are not filled when the emission is arising from the back side of the infalling shell. The contour of the slow wide-angle component is indicated by the white dashed line. The color scale is the same for all the channel maps.}
    \label{fig:OuterOutflow}
\end{figure*}

\SCC{Our ALMA maps reveal that} the inner conical outflow is nested inside a  wider component ($R \approx 10^{\prime\prime}$ at $Z = 15^{\prime\prime}$)  observed at low redshifted velocities ($V-V_{\rm sys} = 1-2$ \kms).  This structure is shown in Figs. \ref{fig:Moneyplot} and \ref{fig:OuterOutflow} and can also be seen
on transverse PV diagrams in the form of an extended pedestal with a shallower $V(r)$ slope at $(V-V_{\rm sys})   \leq 2$~\kms~(\al{red} arrows on Fig.~\ref{fig:XVcone}). An important part of the $^{12}$CO emission from this component falls at velocities $(V-V_{ \rm sys}) \le 1$ \kms\ where it is absorbed by the surrounding medium or resolved out by ALMA. Contrary to the conical outflow, the wide and slow component shows a parabolic morphology \COI{that shifts away from the source along the outflow axis as the line-of-sight velocity increases}
for ($V-V_{sys}$) = 1.2 to 1.8 \kms~(see Fig.~\ref{fig:OuterOutflow}).
Discrete loops are also observed in the channel maps, suggesting time variability and  contributions from different layers \SCC{inside} \al{this region}. A strong transverse asymmetry of the redshifted emission can be observed, for instance, at $(V-V_{\rm sys})= +1.2$~\kms, 
\SCC{namely; the emission at a given height is more extended toward $\delta x > 0$ than $\delta x < 0$. Transverse PV cuts in Fig.~\ref{fig:XVcone} show that this is due to a global "skew" across the low-velocity pedestal, in the sense that emission at positive \qu{$\delta x$} is shifted slightly further into the red compared to symmetric negative \qu{$\delta x$} where it is strongly absorbed. This skew is}
indicative of rotation in the same direction as the disk (and the inner conical flow), \SCC{but because of absorption by the ambient cloud, it cannot be easily quantified.}

In Fig.~\ref{fig:OuterOutflow}, we compare the morphology and kinematics of this wide-angle component with analytical predictions for an axisymetric rotating shell falling ballistically \SC{onto \COI{the northwest surface of }the disk} \COI{around a central point mass of 1\al{.1}~\Msun} using the dynamics and structure computed by \citet{ulrich_infall_1976} (see details in Appendix \ref{sec:Ulrich}). \SC{Although the model is idealized, it allows one to visualize the key 3D projection effects expected at $i = 63\degr$ in such a geometry}. The predicted locus of the shell in each channel map is shown as white contours, for a centrifugal radius $R_d$=700~au and an initial polar angle from the disk axis $\theta_0$=70$^{\circ}$. The effect of varying $R_d$ 
and $\theta_0$ is illustrated in Fig.~\ref{fig:Infall}.  No combination of infall parameters $R_d$ and $\theta_0$ can reproduce both the large extent and the apparent acceleration of the wide redshifted component. The match would be even worse if infall is slown down by pressure gradients, as in the Class I source TMC1A \citep{aso_alma_2015}. The only feature indicative of infall in the receding lobe of the system is the blueshifted arc opening toward \COI{the} north\COI{west} from $z \simeq -2.2''$, which is suggestive of the back side of an infalling shell from  $\theta_0 > 70\degr$ (see leftmost panel of Fig.~\ref{fig:OuterOutflow}).  Therefore, we conclude that residual infall in the receding lobe is confined to  equatorial regions, and the wide angle redshifted component surrounding the inner cone traces mostly outflowing material.

\section{Discussion}
\label{Discussion}

\subsection{\al{DG Tau B disk: Evolutionary stage} and origin of structure}
\label{sec:structure_disk}

DG~Tau~B has been traditionally classified as a Class~I source based on the slope of its near-infrared spectral energy distribution \citep{luhman_disk_2010}. However, \citet{crapsi_characterizing_2008} showed that strongly inclined Class II~disks (i $\ge$ 65 $^\circ$) could be misclassified as Class~I with the spectral slope method, and \citet{kruger_gas_2011} attribute ice absorptions toward DG~Tau~B to a thick edge-on disk. Our ALMA observations show that while DG Tau B's disk is indeed very large and flared, residual infall onto the disk is still present 
at large polar angles \SCC{$\theta_0 >$ 70\degr} in both lobes of the system (see Sects. 3.1 and 3.2.2). However, by extrapolating the single-dish fluxes obtained by Mottram et al. (2017) at 450$\mu$m and 850$\mu$m, we find that the extended dust envelope around DG Tau B contributes only $300 _{-300}^{+400}$~mJy at 1.3mm, that is, at most comparable to the disk \LEt{flux} itself. Following Eq. 1 of \citet{motte_circumstellar_2001}, we would infer an envelope mass < 0.37 M$_{\odot}$ within 15$^{\prime\prime}$ (2000 au), that is,
less than one third the stellar mass. Hence we conclude that DG Tau B is in the late Class I phase, \SCC{and it is \COI{in the final stages of clearing up}} its infalling envelope. 

We note that DG Tau B's disk is as large as the largest known Class II disks. Two other large Class I disks with dust and gas sizes similar to DG Tau B are GY 91 \citep[\RDUST~=~140 au,][]{van_der_marel_protoplanetary_2019}
and L1489-IRS \citep[\RDUST~=~250 au, \RCO~=~700 au,][]{yen_alma_2014}. Hence, the trend of increasing gas disk size with age reported by \citet{najita_protoplanetary_2018} may be affected by low-number statistics, and an unbiased survey of Class I disks appears necessary to investigate disk size evolution at $\le 1$ Myr.
 
\al{DG tau B is the third Class I \COI{disk} where \qu{axisymmetric bumps of enhanced emission} were reported, after TMC1-A 
\citep{bjerkeli_resolved_2016,harsono_evidence_2018} and GY~91 \citep{sheehan_multiple_2018}}.
\qu{All three \COI{disks} show bumps and dips with low contrast compared to the main emission profile.}


\al{In the model continuum intensity profile \SCC{of DG Tau B}, we see a depression preceding the first emission bump (Fig.~\ref{fig:profil_sconv})} \SCC{that might} trace the possible signature of a gap. \al{Following the method outlined in \citet{long_gaps_2018}, we estimate a gap center of $R_g$=53~au with a width of $\Delta$=9~au,} which is typical of those recently observed in Class~II disks \citep{long_gaps_2018,andrews_disk_2018}. 
Assuming that $\Delta$ is 5.5 times the planet's Hill radius \citep{lodato_newborn_2019}, the mass of the planet that could carve the inner gap in DG~Tau~B would be 0.1 M$_{\rm J}$.

\al{Alternatively, the observed emission bumps in the DG~Tau~B disk may be due to other mechanisms than rings and gaps carved by planets. Indeed, the radial emission profile does not show \SCC{dips going} below the underlying \SCC{centrally peaked power-law} component, \SCC{which is unlike what has been} observed, for example, in the HL~Tau disk \citep{brogan_2014_2015}. We might therefore be witnessing an earlier stage of millimetric dust accumulation before planetary and gap formation. Different mechanisms can lead to the formation of radial structures in disks.}
Dust radial trapping and growth due, for example, to snow lines \citep{Zhang15} as well as the secular gravitational instability \citep{takahashi_origin_2016} have been recently invoked to reproduce the rings in the HL Tau disk. This latter mechanism could act in the large and likely massive DG Tau B disk. 
\al{Nonideal magnetohydrodynamic (MHD) effects and zonal flows can also generate radial structures in disks, with lower contrast at large radii \citep[e.g., in][]{bethune_global_2017,suriano_formation_2019}.}
\qu{In any case, our DG~Tau~B ALMA observations confirm that disk substructures can form at an early stage where infall and outflow are still present and could potentially affect the disk structure and possibly planet formation.}

\subsection{Origin of the nested rotating outflows in DG~Tau~B}
\label{section:WDS}

\al{Our ALMA observations reveal that the inner bright conical CO outflow in DG Tau B is surrounded by outflowing material. Therefore, it is not tracing the interface between infall and outflow}. This interface must be located at much larger angles ($\ge 70$\degr~from the geometry of the lowest velocity shells observed in our channel maps, \SCC{compared with infall models, see Figs. 3 and 4)}. 
\SCC{An important implication of} these results \SCC{is} that outflowing material can occupy a wider solid angle than scattered light cavities suggest. 
This may explain why classical disk plus envelope models have difficulties in reproducing the spectral energy distribution and scattered light cavities images in DG~Tau~B at the same time \citep{gramajo_combined_2010} as well as why some Class II sources with active outflows appear to have dusty conical "envelopes" above their disks \citep[e.g., RY~Tau, \al{Fig.~3} in][]{takami_high-contrast_2013}

The parabolic morphology of the slow wide angle flow and its multiple inner loops resemble the structure of slow outflow cavities around younger sources, such as HH46-47 \citep{zhang_episodic_2019}. They are interpreted as ambient material swept up by an episodic wide-angle inner wind. \al{Wind-driven shell models predict emission maps \COI{shifting away from the star as the line-of-sight velocity increases }as observed in this component \citep[see Fig.~24 in ][]{lee_co_2000} }. 

\qu{However,} the striking conical inner flow \al{anchored at small radii $\le $10--60 au} highly \qu{unusual so far. It is} reminiscent of the predictions for the long-term interaction shell between inner jet bowshocks and a steady outer disk wind, as recently computed by \citet{tabone_interaction_2018}. \al{The interaction between an outer disk wind and an inner \SCC{wide-angle wind} cannot be excluded either, but \COI{predictions for this scenario do not exist yet.}}

\qu{Alternatively, the constant velocity and mass flux of the inner conical flow, also seen in HH30 \citep{louvet_hh30_2018}, suggest that it could trace  matter directly ejected from the disk.}
\MVSC{In the case of \SCC{a cold, steady, axisymmetric} MHD disk wind, our derived value of angular momentum and poloidal speed would predict an ejection radius of r$_{0}$ $\approx$ 2~au and a magnetic level arm of $\lambda = 1.6$, \SCC{using the formulae in \citet{anderson_locating_2003}}.}
This model was ruled out by \citet{zapata_kinematics_2015} based on the  small magnetic lever arm $\lambda \simeq \al{2}$ and large outflow mass-flux. However, recent nonideal MHD simulations of magnetized winds from protoplanetary disks do predict small magnetic lever arms and a mass-flux of the same order as the accretion rate \citep{bethune_global_2017,bai_magneto-thermal_2016}, as observed here. 
\qu{They also predict an important role of thermal gradients in driving the wind. In that case, the launching radius $r_0$ could be up to 5 au (corresponding to the minimum $ \lambda \approx 1$) }.


\al{However, we stress that our estimate of angular momentum is highly uncertain given the complex} \al{steep velocity gradients}, projection effects, and \SCC{beam smearing effects.} Three-dimensional modeling of \SCC{the observations} is requested to \SCC{constrain the flow rotation and kinematic structure more accurately as well as to firmly} distinguish between the various scenarios. This is outside the scope of the present Letter and will be the subject of a \SCC{forthcoming} paper \SCC{(de Valon et al., in prep)}. 

Whatever its origin, we stress that the large mass-flux carried along the inner conical outflow walls, being comparable to the inner disk accretion rate, must play a major role in limiting the final mass of the star and disk. 
Thanks to its large disk, proximity, and favorable inclination (compared e.g., to HH30),
DG~Tau~B is a very promising target to elucidate the disk-outflow connexion and its impact on disk evolution and planet formation.

\begin{acknowledgements}
\qu{The authors would like to thank the referee, whose comments helped improve the quality of the paper. }We acknowledge Ana Lopez-Sepulcre for crucial help in the data reduction process at the IRAM-Grenoble ALMA ARC node. This paper makes use of the following ALMA data: ADS/JAO.ALMA\#2015.1.01108.S, ADS/JAO.ALMA\#2017.1.01605.S. ALMA is a partnership of ESO (representing its member states), NSF (USA) and NINS (Japan), together with NRC (Canada), MOST and ASIAA (Taiwan), and KASI (Republic of Korea), in cooperation with the Republic of Chile.The Joint ALMA Observatory is operated by ESO, AUI/NRAO and NAOJ. Part of the data reduction presented in this paper were performed on the GRICAD infrastructure (https://gricad.univ-grenoble-alpes.fr), which is partly supported by the Equipe@Meso project (reference ANR-10-EQPX-29-01) of the programme Investissements d'Avenir supervised by the Agence Nationale pour la Recherche. This work was supported by the Programme National de Physique Stellaire (PNPS) 
and the Programme National de Physique et Chimie du Milieu Interstellaire (PCMI) 
of CNRS/INSU co-funded by CEA and CNES. L.A.Z acknowledge financial support from DGAPA, UNAM, and CONACyT, M\'exico. F.L acknowledges the support of the Fondecyt program n$_{\circ}$ 3170360.
\end{acknowledgements}
\bibliographystyle{aa} 
\bibliography{Lib_letter} 

\begin{thebibliography}{47}
\expandafter\ifx\csname natexlab\endcsname\relax\def\natexlab#1{#1}\fi

\bibitem[{Anderson {et~al.}(2003)Anderson, Li, Krasnopolsky, \&
  Blandford}]{anderson_locating_2003}
Anderson, J.~M., Li, Z.-Y., Krasnopolsky, R., \& Blandford, R.~D. 2003, The
  Astrophysical Journal, 590, L107

\bibitem[{Andrews {et~al.}(2018)Andrews, Huang, P\'erez, Isella, Dullemond,
  Kurtovic, GuzmÃ¡n, Carpenter, Wilner, Zhang, Zhu, Birnstiel, Bai, Benisty,
  Hughes, Ãberg, \& Ricci}]{andrews_disk_2018}
Andrews, S.~M., Huang, J., P\'erez, L.~M., {et~al.} 2018, The Astrophysical
  Journal, 869, L41

\bibitem[{Aso {et~al.}(2015)Aso, Ohashi, Saigo, Koyamatsu, Aikawa, Hayashi,
  Machida, Saito, Takakuwa, Tomida, Tomisaka, \& Yen}]{aso_alma_2015}
Aso, Y., Ohashi, N., Saigo, K., {et~al.} 2015, The Astrophysical Journal, 812,
  27

\bibitem[{Bai {et~al.}(2016)Bai, Ye, Goodman, \&
  Yuan}]{bai_magneto-thermal_2016}
Bai, X.-N., Ye, J., Goodman, J., \& Yuan, F. 2016, The Astrophysical Journal,
  818, 152

\bibitem[{B\'ethune {et~al.}(2017)B\'ethune, Lesur, \&
  Ferreira}]{bethune_global_2017}
B\'ethune, W., Lesur, G., \& Ferreira, J. 2017, Astronomy \& Astrophysics, 600,
  A75

\bibitem[{Bjerkeli {et~al.}(2016)Bjerkeli, van~der Wiel, Harsono, Ramsey, \&
  J{\o}rgensen}]{bjerkeli_resolved_2016}
Bjerkeli, P., van~der Wiel, M. H.~D., Harsono, D., Ramsey, J.~P., \&
  J{\o}rgensen, J.~K. 2016, Nature, 540, 406

\bibitem[{Brogan {et~al.}(2015)Brogan, Pérez, Hunter, Dent, Hales, Hills,
  Corder, Fomalont, Vlahakis, Asaki, Barkats, Hirota, Hodge, Impellizzeri,
  Kneissl, Liuzzo, Lucas, Marcelino, Matsushita, Nakanishi, Phillips, Richards,
  Toledo, Aladro, Broguiere, Cortes, Cortes, Espada, Galarza, Appadoo, Ramirez,
  Humphreys, Jung, Kameno, Laing, Leon, Marconi, Mignano, Nikolic, Nyman,
  Radiszcz, Remijan, Rodón, Sawada, Takahashi, Tilanus, Vilaro, Watson,
  Wiklind, Akiyama, Chapillon, Monsalvo, Francesco, Gueth, Kawamura, Lee,
  Luong, Mangum, Pietu, Sanhueza, Saigo, Takakuwa, Ubach, Kempen, Wootten,
  Carrizo, Francke, Gallardo, Garcia, Gonzalez, Hill, Kaminski, Kurono, Liu,
  Lopez, Morales, Plarre, Schieven, Testi, Videla, Villard, Andreani, Hibbard,
  \& Tatematsu}]{brogan_2014_2015}
Brogan, C.~L., Pérez, L.~M., Hunter, T.~R., {et~al.} 2015, The Astrophysical
  Journal, 808, L3

\bibitem[{Brown {et~al.}(2018)Brown, Vallenari, Prusti, Bruijne, Babusiaux,
  Bailer-Jones, Biermann, Evans, Eyer, Jansen, Jordi, Klioner, Lammers,
  Lindegren, Luri, Mignard, Panem, Pourbaix, Randich, Sartoretti, Siddiqui,
  Soubiran, Leeuwen, Walton, Arenou, Bastian, Cropper, Drimmel, Katz, Lattanzi,
  Bakker, Cacciari, CastaÃ±eda, Chaoul, Cheek, Angeli, Fabricius, Guerra,
  Holl, Masana, Messineo, Mowlavi, Nienartowicz, Panuzzo, Portell, Riello,
  Seabroke, Tanga, ThÃ©venin, Gracia-Abril, Comoretto, Garcia-Reinaldos,
  Teyssier, Altmann, Andrae, Audard, Bellas-Velidis, Benson, Berthier, Blomme,
  Burgess, Busso, Carry, Cellino, Clementini, Clotet, Creevey, Davidson,
  Ridder, Delchambre, DellâOro, Ducourant, FernÃ¡ndez-HernÃ¡ndez,
  Fouesneau, FrÃ©mat, Galluccio, GarcÃ­a-Torres, GonzÃ¡lez-NÃºÃ±ez,
  GonzÃ¡lez-Vidal, Gosset, Guy, Halbwachs, Hambly, Harrison, HernÃ¡ndez,
  Hestroffer, Hodgkin, Hutton, Jasniewicz, Jean-Antoine-Piccolo, Jordan, Korn,
  Krone-Martins, Lanzafame, Lebzelter, LÃ¶ffler, Manteiga, Marrese,
  MartÃ­n-Fleitas, Moitinho, Mora, Muinonen, Osinde, Pancino, Pauwels, Petit,
  Recio-Blanco, Richards, Rimoldini, Robin, Sarro, Siopis, Smith, Sozzetti,
  SÃ¼veges, Torra, Reeven, Abbas, Aramburu, Accart, Aerts, Altavilla,
  Ãlvarez, Alvarez, Alves, Anderson, Andrei, Varela, Antiche, Antoja, Arcay,
  Astraatmadja, Bach, Baker, Balaguer-NÃºÃ±ez, Balm, Barache, Barata,
  Barbato, Barblan, Barklem, Barrado, Barros, Barstow, MuÃ±oz, Bassilana,
  Becciani, Bellazzini, Berihuete, Bertone, Bianchi, BienaymÃ©,
  Blanco-Cuaresma, Boch, Boeche, Bombrun, Borrachero, Bossini, Bouquillon,
  Bourda, Bragaglia, Bramante, Breddels, Bressan, Brouillet, BrÃ¼semeister,
  Brugaletta, Bucciarelli, Burlacu, Busonero, Butkevich, Buzzi, Caffau,
  Cancelliere, Cannizzaro, Cantat-Gaudin, Carballo, Carlucci, Carrasco,
  Casamiquela, Castellani, Castro-Ginard, Charlot, Chemin, Chiavassa, Cocozza,
  Costigan, Cowell, Crifo, Crosta, Crowley, Cuypersâ , Dafonte, Damerdji,
  Dapergolas, David, David, Laverny, Luise, March, Martino, Souza, Torres,
  Debosscher, Pozo, Delbo, Delgado, Delgado, Matteo, Diakite, Diener,
  Distefano, Dolding, Drazinos, DurÃ¡n, Edvardsson, Enke, Eriksson, Esquej,
  Bontemps, Fabre, Fabrizio, Faigler, FalcÃ£o, Casas, Federici, Fedorets,
  Fernique, Figueras, Filippi, Findeisen, Fonti, Fraile, Fraser, FrÃ©zouls,
  Gai, Galleti, Garabato, GarcÃ­a-Sedano, Garofalo, Garralda, Gavel, Gavras,
  Gerssen, Geyer, Giacobbe, Gilmore, Girona, Giuffrida, Glass, Gomes, Granvik,
  Gueguen, Guerrier, Guiraud, GutiÃ©rrez-SÃ¡nchez, Haigron, Hatzidimitriou,
  Hauser, Haywood, Heiter, Helmi, Heu, Hilger, Hobbs, Hofmann, Holland, Huckle,
  Hypki, Icardi, JanÃen, Fombelle, Jonker, JuhÃ¡sz, Julbe, Karampelas,
  Kewley, Klar, Kochoska, Kohley, Kolenberg, Kontizas, Kontizas, Koposov,
  Kordopatis, Kostrzewa-Rutkowska, Koubsky, Lambert, Lanza, Lasne, Lavigne,
  Fustec, Poncin-Lafitte, Lebreton, Leccia, Leclerc, Lecoeur-Taibi, Lenhardt,
  Leroux, Liao, Licata, LindstrÃ¸m, Lister, Livanou, Lobel, LÃ³pez,
  Managau, Mann, Mantelet, Marchal, Marchant, Marconi, Marinoni, MarschalkÃ³,
  Marshall, Martino, Marton, Mary, Massari, MatijeviÄ, Mazeh, McMillan,
  Messina, Michalik, Millar, Molina, Molinaro, MolnÃ¡r, Montegriffo, Mor,
  Morbidelli, Morel, Morris, Mulone, Muraveva, Musella, Nelemans, Nicastro,
  Noval, OâMullane, OrdÃ©novic, OrdÃ³Ã±ez-Blanco, Osborne, Pagani,
  Pagano, Pailler, Palacin, Palaversa, Panahi, Pawlak, Piersimoni, Pineau,
  Plachy, Plum, Poggio, Poujoulet, PrÅ¡a, Pulone, Racero, Ragaini, Rambaux,
  Ramos-Lerate, Regibo, ReylÃ©, Riclet, Ripepi, Riva, Rivard, Rixon,
  Roegiers, Roelens, Romero-GÃ³mez, Rowell, Royer, Ruiz-Dern, Sadowski,
  SellÃ©s, Sahlmann, Salgado, Salguero, Sanna, Santana-Ros, Sarasso,
  Savietto, Schultheis, Sciacca, Segol, Segovia, SÃ©gransan, Shih, Siltala,
  Silva, Smart, Smith, Solano, Solitro, Sordo, Nieto, Souchay, Spagna, Spoto,
  Stampa, Steele, SteidelmÃ¼ller, Stephenson, Stoev, Suess, Surdej, Szabados,
  Szegedi-Elek, Tapiador, Taris, Tauran, Taylor, Teixeira, Terrett,
  Teyssandier, Thuillot, Titarenko, Clotet, Turon, Ulla, Utrilla, Uzzi,
  Vaillant, Valentini, Valette, Elteren, Hemelryck, Leeuwen, Vaschetto,
  Vecchiato, Veljanoski, Viala, Vicente, Vogt, Essen, Voss, Votruba, Voutsinas,
  Walmsley, Weiler, Wertz, Wevers, Wyrzykowski, Yoldas, Å½erjal, Ziaeepour,
  Zorec, Zschocke, Zucker, Zurbach, \& Zwitter}]{brown_gaia_2018}
Brown, A. G.~A., Vallenari, A., Prusti, T., {et~al.} 2018, Astronomy \&
  Astrophysics, 616, A1

\bibitem[{Crapsi {et~al.}(2008)Crapsi, van Dishoeck, Hogerheijde, Pontoppidan,
  \& Dullemond}]{crapsi_characterizing_2008}
Crapsi, A., van Dishoeck, E.~F., Hogerheijde, M.~R., Pontoppidan, K.~M., \&
  Dullemond, C.~P. 2008, Astronomy and Astrophysics, 486, 245

\bibitem[{Frank {et~al.}(2014)Frank, Ray, Cabrit, Hartigan, Arce, Bacciotti,
  Bally, Benisty, Eislöffel, Güdel, Lebedev, Nisini, \&
  Raga}]{frank_jets_2014}
Frank, A., Ray, T.~P., Cabrit, S., {et~al.} 2014, Protostars and Planets VI,
  451

\bibitem[{Goldsmith {et~al.}(2008)Goldsmith, Heyer, Narayanan, Snell, Li, \&
  Brunt}]{goldsmith_largescale_2008}
Goldsmith, P.~F., Heyer, M., Narayanan, G., {et~al.} 2008, The Astrophysical
  Journal, 680, 428

\bibitem[{Gramajo {et~al.}(2010)Gramajo, Whitney, G\'omez, \&
  Robitaille}]{gramajo_combined_2010}
Gramajo, L.~V., Whitney, B.~A., G\'omez, M., \& Robitaille, T.~P. 2010, The
  Astronomical Journal, 139, 2504

\bibitem[{Guilloteau {et~al.}(2011)Guilloteau, Dutrey, Pi\'etu, \&
  Boehler}]{guilloteau_dual-frequency_2011}
Guilloteau, S., Dutrey, A., Pi\'etu, V., \& Boehler, Y. 2011, Astronomy \&
  Astrophysics, 529, A105

\bibitem[{Harsono {et~al.}(2018)Harsono, Bjerkeli, van~der Wiel, Ramsey, Maud,
  Kristensen, \& Jørgensen}]{harsono_evidence_2018}
Harsono, D., Bjerkeli, P., van~der Wiel, M. H.~D., {et~al.} 2018, Nature
  Astronomy, 2, 646

\bibitem[{Hartigan {et~al.}(1994)Hartigan, Morse, \&
  Raymond}]{hartigan_mass-loss_1994}
Hartigan, P., Morse, J.~A., \& Raymond, J. 1994, The Astrophysical Journal,
  436, 125

\bibitem[{Kruger {et~al.}(2011)Kruger, Richter, Carr, Najita, Doppmann, \&
  Seifahrt}]{kruger_gas_2011}
Kruger, A.~J., Richter, M.~J., Carr, J.~S., {et~al.} 2011, The Astrophysical
  Journal, 729, 145

\bibitem[{Lee {et~al.}(2000)Lee, Mundy, Reipurth, Ostriker, \&
  Stone}]{lee_co_2000}
Lee, C.-F., Mundy, L.~G., Reipurth, B., Ostriker, E.~C., \& Stone, J.~M. 2000,
  The Astrophysical Journal, 542, 925

\bibitem[{Lodato {et~al.}(2019)Lodato, Dipierro, Ragusa, Long, Herczeg,
  Pascucci, Pinilla, Manara, Tazzari, Liu, Mulders, Harsono, Boehler, Menard,
  Johnstone, Salyk, van~der Plas, Cabrit, Edwards, Fischer, Hendler, Nisini,
  Rigliaco, Avenhaus, Banzatti, \& Gully-Santiago}]{lodato_newborn_2019}
Lodato, G., Dipierro, G., Ragusa, E., {et~al.} 2019, Monthly Notices of the
  Royal Astronomical Society, 486, 453

\bibitem[{Long {et~al.}(2018)Long, Pinilla, Herczeg, Harsono, Dipierro,
  Pascucci, Hendler, Tazzari, Ragusa, Salyk, Edwards, Lodato, van~de Plas,
  Johnstone, Liu, Boehler, Cabrit, Manara, Menard, Mulders, Nisini, Fischer,
  Rigliaco, Banzatti, Avenhaus, \& Gully-Santiago}]{long_gaps_2018}
Long, F., Pinilla, P., Herczeg, G.~J., {et~al.} 2018, The Astrophysical
  Journal, 869, 17

\bibitem[{Louvet {et~al.}(2018)Louvet, Dougados, Cabrit, Mardones, M\'enard,
  Tabone, Pinte, \& Dent}]{louvet_hh30_2018}
Louvet, F., Dougados, C., Cabrit, S., {et~al.} 2018, Astronomy \& Astrophysics,
  618, A120

\bibitem[{Luhman {et~al.}(2010)Luhman, Allen, Espaillat, Hartmann, \&
  Calvet}]{luhman_disk_2010}
Luhman, K.~L., Allen, P.~R., Espaillat, C., Hartmann, L., \& Calvet, N. 2010,
  The Astrophysical Journal Supplement Series, 186, 111

\bibitem[{McMullin {et~al.}(2007)McMullin, Waters, Schiebel, Young, \&
  Golap}]{mcmullin_casa_2007}
McMullin, J.~P., Waters, B., Schiebel, D., Young, W., \& Golap, K. 2007,
  Astronomical Data Analysis Software and Systems XVI, 376, 127

\bibitem[{Mitchell {et~al.}(1997)Mitchell, Sargent, \&
  Mannings}]{mitchell_dg_1997}
Mitchell, G.~F., Sargent, A.~I., \& Mannings, V. 1997, The Astrophysical
  Journal, 483, L127

\bibitem[{Motte \& André(2001)}]{motte_circumstellar_2001}
Motte, F. \& André, P. 2001, Astronomy \& Astrophysics, 365, 440

\bibitem[{Mottram {et~al.}(2017)Mottram, van Dishoeck, Kristensen, Karska, San
  JosÃ©-GarcÃ­a, Khanna, Herczeg, AndrÃ©, Bontemps, Cabrit, Carney,
  Drozdovskaya, Dunham, Evans, Fedele, Green, Harsono, Johnstone, JÃ¸rgensen,
  KÃ¶nyves, Nisini, Persson, Tafalla, Visser, \&
  YÄ±ldÄ±z}]{mottram_outflows_2017}
Mottram, J.~C., van Dishoeck, E.~F., Kristensen, L.~E., {et~al.} 2017,
  Astronomy \& Astrophysics, 600, A99

\bibitem[{Mundt \& Fried(1983)}]{mundt_jets_1983}
Mundt, R. \& Fried, J.~W. 1983, The Astrophysical Journal, 274, L83

\bibitem[{Mundt {et~al.}(1991)Mundt, Ray, \& Raga}]{mundt_collimation_1991}
Mundt, R., Ray, T.~P., \& Raga, A.~C. 1991, Astronomy and Astrophysics, 252,
  740

\bibitem[{Najita \& Bergin(2018)}]{najita_protoplanetary_2018}
Najita, J.~R. \& Bergin, E.~A. 2018, The Astrophysical Journal, 864, 168

\bibitem[{Padgett {et~al.}(1999)Padgett, Brandner, Stapelfeldt, Strom, Terebey,
  \& Koerner}]{padgett_hst/nicmos_1999}
Padgett, D.~L., Brandner, W., Stapelfeldt, K.~R., {et~al.} 1999, The
  Astronomical Journal, 117, 1490

\bibitem[{Pinte {et~al.}(2018)Pinte, M\'enard, Duch\^ene, Hill, Dent, Woitke,
  Maret, van~der Plas, Hales, Kamp, Thi, de~Gregorio-Monsalvo, Rab, Quanz,
  Avenhaus, Carmona, \& Casassus}]{pinte_direct_2018}
Pinte, C., M\'enard, F., Duch\^ene, G., {et~al.} 2018, Astronomy \&
  Astrophysics, 609, A47

\bibitem[{Podio {et~al.}(2011)Podio, Eisl\"offel, Melnikov, Hodapp, \&
  Bacciotti}]{podio_tracing_2011}
Podio, L., Eisl\"offel, J., Melnikov, S., Hodapp, K.~W., \& Bacciotti, F. 2011,
  Astronomy \& Astrophysics, 527, A13

\bibitem[{Schmalzl {et~al.}(2010)Schmalzl, Kainulainen, Quanz, Alves, Goodman,
  Henning, Launhardt, Pineda, \& RomÃ¡n-ZÃºÃ±iga}]{schmalzl_star_2010}
Schmalzl, M., Kainulainen, J., Quanz, S.~P., {et~al.} 2010, The Astrophysical
  Journal, 725, 1327

\bibitem[{Sheehan \& Eisner(2018)}]{sheehan_multiple_2018}
Sheehan, P.~D. \& Eisner, J.~A. 2018, The Astrophysical Journal, 857, 18

\bibitem[{Suriano {et~al.}(2019)Suriano, Li, Krasnopolsky, Suzuki, \&
  Shang}]{suriano_formation_2019}
Suriano, S.~S., Li, Z.-Y., Krasnopolsky, R., Suzuki, T.~K., \& Shang, H. 2019,
  Monthly Notices of the Royal Astronomical Society, 484, 107

\bibitem[{Tabone {et~al.}(2018)Tabone, Raga, Cabrit, \& Pineau~des
  For\^ets}]{tabone_interaction_2018}
Tabone, B., Raga, A., Cabrit, S., \& Pineau~des For\^ets, G. 2018, Astronomy \&
  Astrophysics, 614, A119

\bibitem[{Takahashi \& Inutsuka(2016)}]{takahashi_origin_2016}
Takahashi, S.~Z. \& Inutsuka, S.-i. 2016, The Astronomical Journal, 152, 184

\bibitem[{Takami {et~al.}(2013)Takami, Karr, Hashimoto, Kim, Wisniewski,
  Henning, Grady, Kandori, Hodapp, Kudo, Kusakabe, Chou, Itoh, Momose, Mayama,
  Currie, Follette, Kwon, Abe, Brandner, Brandt, Carson, Egner, Feldt, Guyon,
  Hayano, Hayashi, Hayashi, Ishii, Iye, Janson, Knapp, Kuzuhara, McElwain,
  Matsuo, Miyama, Morino, Moro-Martin, Nishimura, Pyo, Serabyn, Suto, Suzuki,
  Takato, Terada, Thalmann, Tomono, Turner, Watanabe, Yamada, Takami, Usuda, \&
  Tamura}]{takami_high-contrast_2013}
Takami, M., Karr, J.~L., Hashimoto, J., {et~al.} 2013, The Astrophysical
  Journal, 772, 145

\bibitem[{Terebey {et~al.}(1984)Terebey, Shu, \&
  Cassen}]{terebey_collapse_1984}
Terebey, S., Shu, F.~H., \& Cassen, P. 1984, The Astrophysical Journal, 286,
  529

\bibitem[{Ulrich(1976)}]{ulrich_infall_1976}
Ulrich, R.~K. 1976, The Astrophysical Journal, 210, 377

\bibitem[{van~der Marel {et~al.}(2019)van~der Marel, Dong, di~Francesco,
  Williams, \& Tobin}]{van_der_marel_protoplanetary_2019}
van~der Marel, N., Dong, R., di~Francesco, J., Williams, J., \& Tobin, J. 2019,
  The Astrophysical Journal, 872, 112

\bibitem[{van Kempen {et~al.}(2009)van Kempen, van Dishoeck, Salter,
  Hogerheijde, JÃ¸rgensen, \& Boogert}]{van_kempen_nature_2009}
van Kempen, T.~A., van Dishoeck, E.~F., Salter, D.~M., {et~al.} 2009, Astronomy
  \& Astrophysics, 498, 167

\bibitem[{Yen {et~al.}(2014)Yen, Takakuwa, Ohashi, Aikawa, Aso, Koyamatsu,
  Machida, Saigo, Saito, Tomida, \& Tomisaka}]{yen_alma_2014}
Yen, H.-W., Takakuwa, S., Ohashi, N., {et~al.} 2014, The Astrophysical Journal,
  793, 1

\bibitem[{Zapata {et~al.}(2015)Zapata, Lizano, Rodr\'iguez, Ho, Loinard,
  Fern\'andez-L\'opez, \& Tafoya}]{zapata_kinematics_2015}
Zapata, L.~A., Lizano, S., Rodr\'iguez, L.~F., {et~al.} 2015, The Astrophysical
  Journal, 798, 131

\bibitem[{{Zhang} {et~al.}(2015){Zhang}, {Blake}, \& {Bergin}}]{Zhang15}
{Zhang}, K., {Blake}, G.~A., \& {Bergin}, E.~A. 2015, \apjl, 806, L7

\bibitem[{Zhang {et~al.}(2016)Zhang, Arce, Mardones, Cabrit, Dunham, Garay,
  Noriega-Crespo, Offner, Raga, \& Corder}]{zhang_alma_2016}
Zhang, Y., Arce, H.~G., Mardones, D., {et~al.} 2016, The Astrophysical Journal,
  832, 158

\bibitem[{Zhang {et~al.}(2019)Zhang, Arce, Mardones, Cabrit, Dunham, Garay,
  Noriega-Crespo, Offner, Raga, \& Corder}]{zhang_episodic_2019}
Zhang, Y., Arce, H.~G., Mardones, D., {et~al.} 2019, The Astrophysical Journal,
  883, 1

\bibitem[{Zhang {et~al.}(2018)Zhang, Higuchi, Sakai, Oya, López-Sepulcre,
  Imai, Sakai, Watanabe, Ceccarelli, Lefloch, \&
  Yamamoto}]{zhang_rotation_2018}
Zhang, Y., Higuchi, A.~E., Sakai, N., {et~al.} 2018, The Astrophysical Journal,
  864, 76

\end{thebibliography}
\newpage

\begin{appendix}
\section{Model of the radial intensity profile }
\label{sec:BestFit}
To model the radial intensity profile of the continuum emission, we used the equation of an exponentially tapered power law \SCC{plus two off-centered Gaussians to reproduce the bumps}:

\noindent
   \begin{equation}
      I(r)=A \Bigg( \frac{r}{r_t} \Bigg)^{\gamma} \exp{\Bigg[ -\Bigg( \frac{r}{r_t} \Bigg)^{\beta} \Bigg]}+\sum_{i} B_i \exp{\Bigg[ -\frac{(r-R_i)^2}{2\sigma_i} \Bigg]}.
      \label{equ:contfit}
   \end{equation}
 
\begin{figure}
    \centering
    \resizebox{\hsize}{!}{\includegraphics{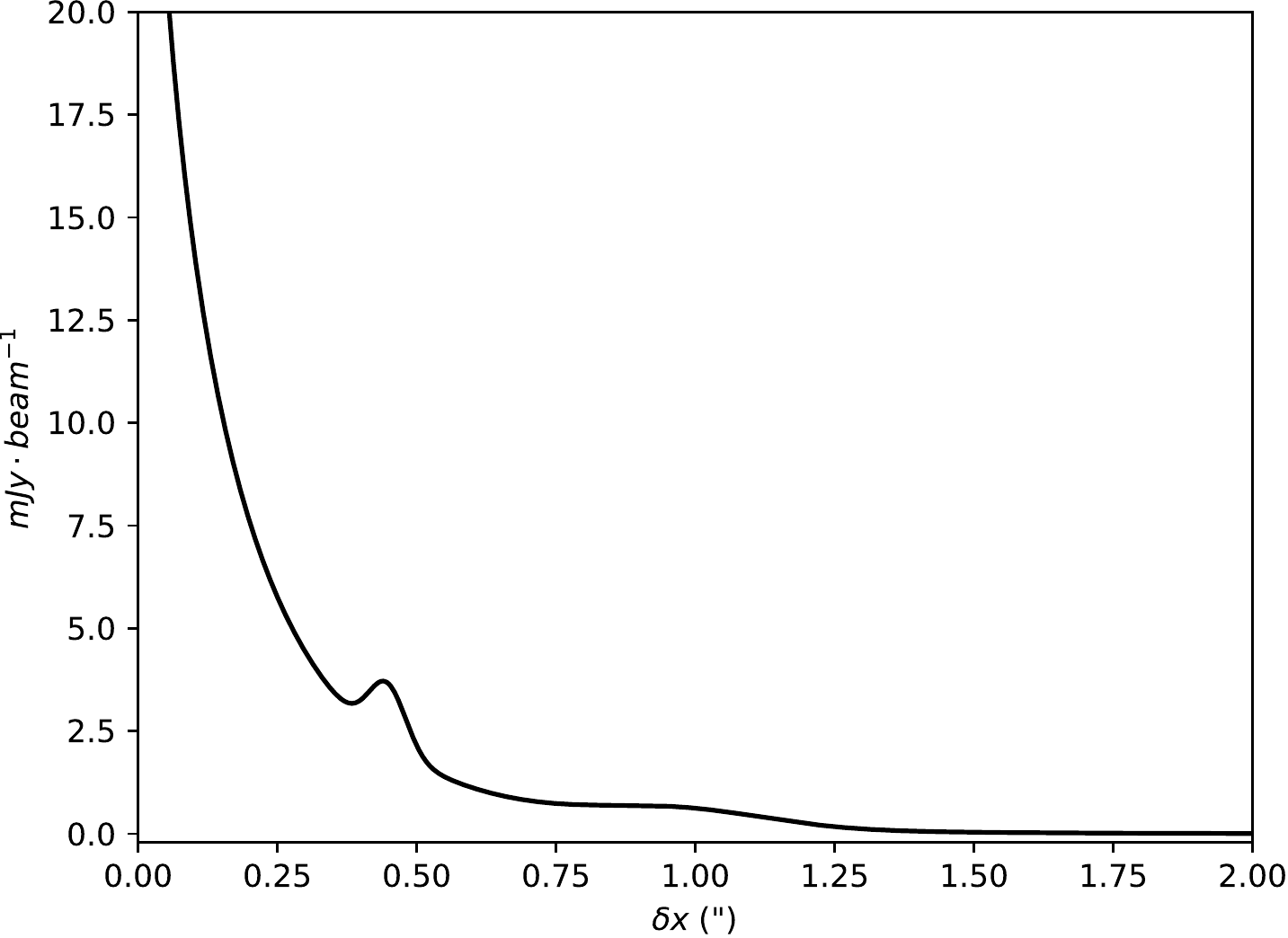}}
    \caption{\al{Best fit to the radial continuum emission profile before convolution by the beam.}}
    \label{fig:profil_sconv}
\end{figure}

To compare effectively with our observations, we convolved this profile with a kernel with a \LEt{full width at half maximum (}FWHM\LEt{)} corresponding to the beam FWHM in that direction. The fitting is achieved using a nonlinear least squares method. The parameters of the best fit with their standard deviation errors are listed in Table \ref{t2} and the model intensity profile without convolution is shown in Fig. \ref{fig:profil_sconv}.

\begin{table*}
\caption{\label{t1} Log of observations}
\centering
\resizebox{\textwidth}{!}{\begin{tabular}{lcccccccc}
\hline\hline
Cycle&Configuration  &Number of &Date& Number of  & Baseline range  & Total on source &Flux & Phase  \\
 & & sessions & (YYYY - MM - DD) & antennas & (m) & integration time & calibrator & calibrator \\
\hline

3 & C36-2/3 & 1 &2016 - 04 - 23  & 39 & 15.1 - 400 & 0h48mn& J0238+1636&J0510+1800\\
3 & C40-5 &  4 (2 kept)\tablefootmark{a}&2016 - 08 - 05/15 & 43 / 38 & 15.1 - 1\,500 / 1\,600 & 1h37mn & J0510+1800&J0510+1800\\
5 & C43-8 &  3 &2017 - 11 - 16/18/19& 43 / 44 / 46 & 92.2 - 8\,500 & 2h06mn & J0510+1800&J0438+3004\\
\hline
\end{tabular}}
\tablefoot{\tablefoottext{a}{Only two sessions were kept due to poor phase calibration.}}
\end{table*}

\section{CO channel maps}

\begin{figure*}
    \centering
    \resizebox{\hsize}{!}{\includegraphics{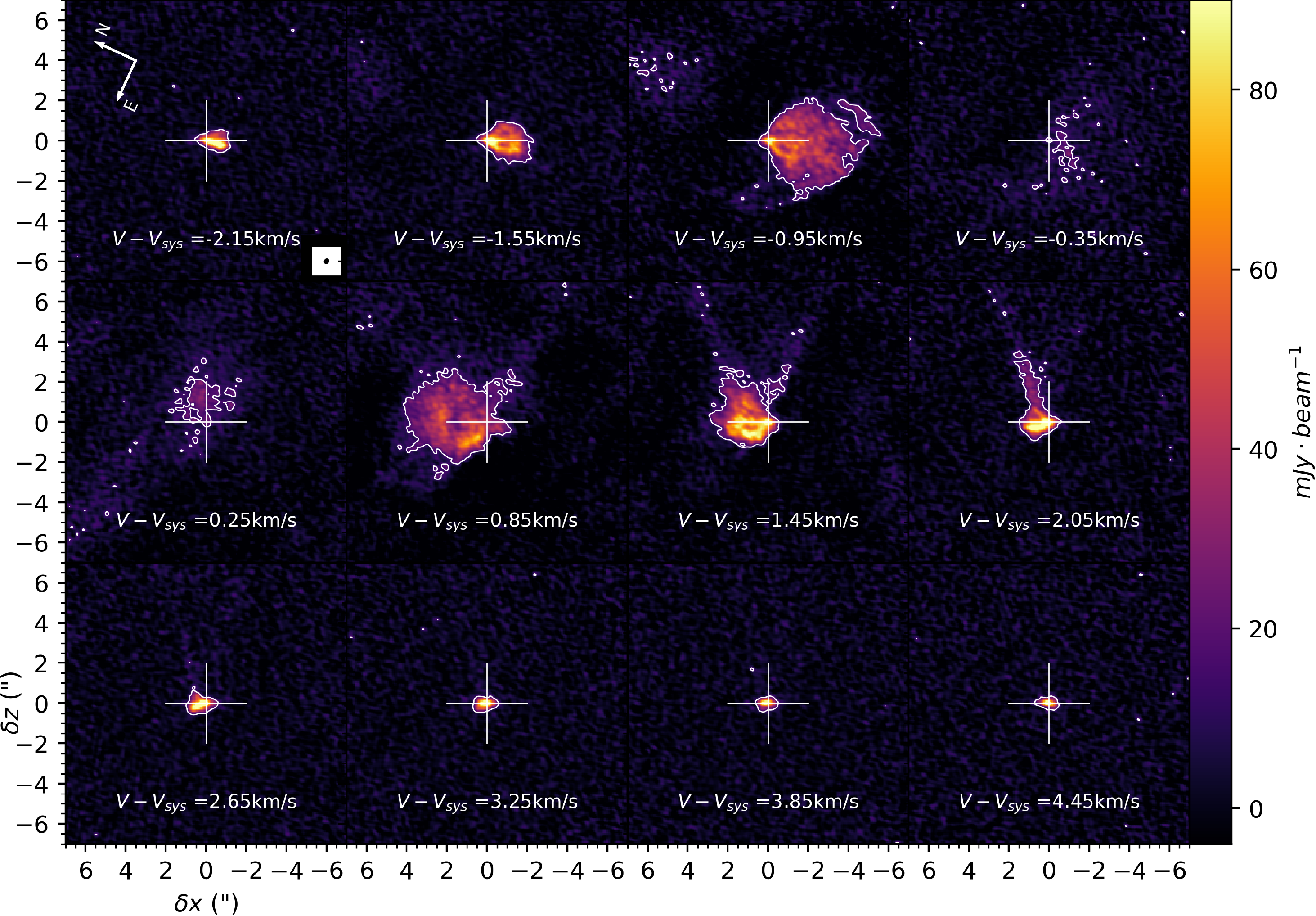}}
    \caption{\al{$^{13}$CO \COI{selected} individual channel maps. For each channel map $\Delta V = 0.3$~\kms~and the central velocity is indicated. The white contours trace the 3$\sigma$ limit with $\sigma$~=~5~mJy/beam (or 1.6~K). The white cross in each panel locates the central position of the continuum. The beam (0.23$^{\prime\prime}$x0.34$^{\prime\prime}$ at P.A. = 0.5$^{\circ}$) is shown in the first panel. }}
    \label{fig:13CO_channels}
\end{figure*}

\begin{figure*}
    \centering
    \resizebox{\hsize}{!}{\includegraphics{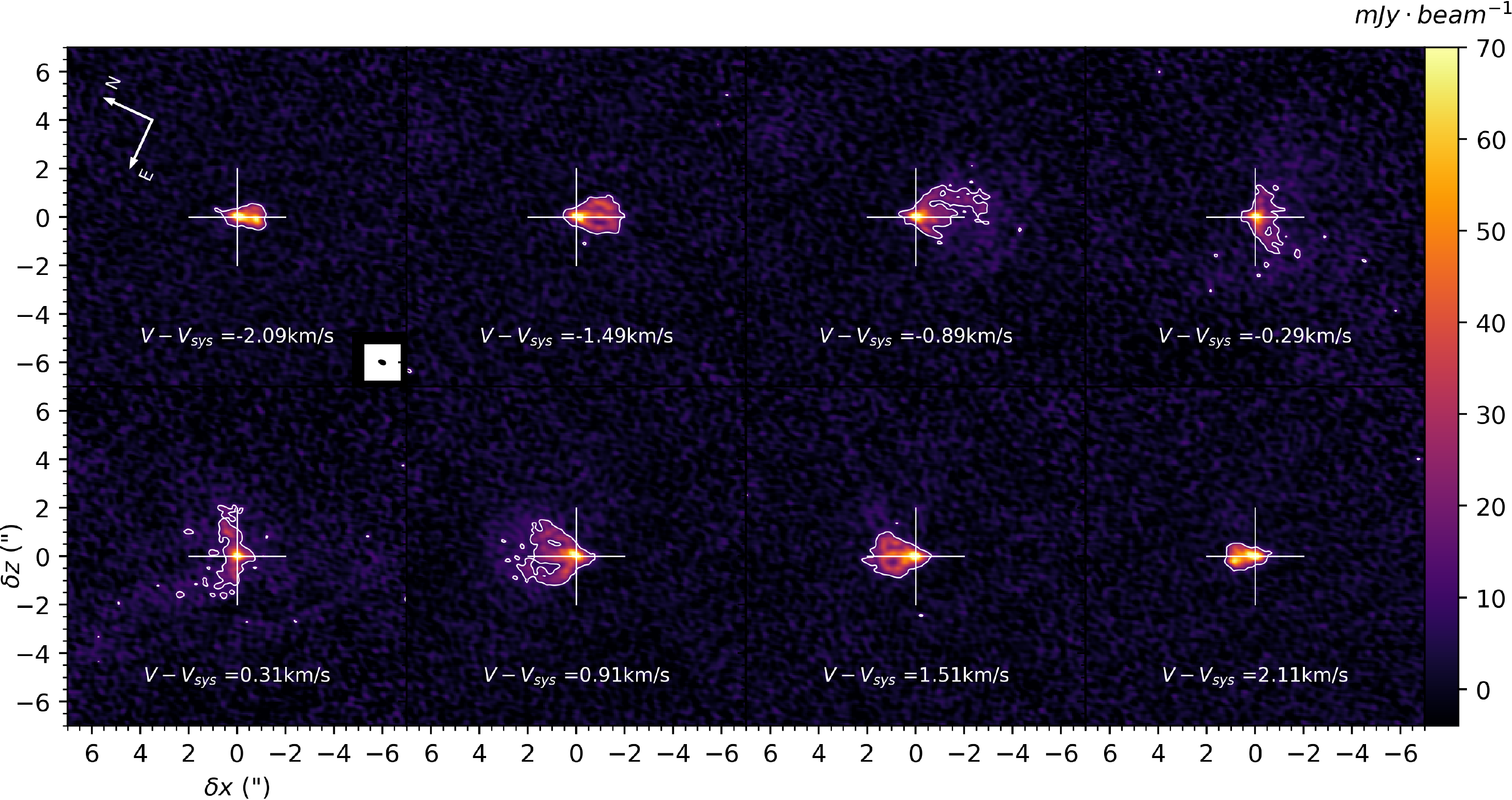}}
    \caption{C$^{18}$O \COI{selected} individual channel maps. \COI{For each channel map  $\Delta V = 0.3$~\kms~and the central velocity is indicated}. The white contours trace the 3$\sigma$ limit with $\sigma$~=~4~mJy/beam (or 1.3~K). The white cross in each panel locates the central position of the continuum. The beam (0.23$^{\prime\prime}$x0.34$^{\prime\prime}$ at P.A. = 2$^{\circ}$) is shown in the first panel.}
    \label{fig:C18O_channels}
\end{figure*}

\begin{figure*}
    \centering
    \resizebox{\hsize}{!}{\includegraphics{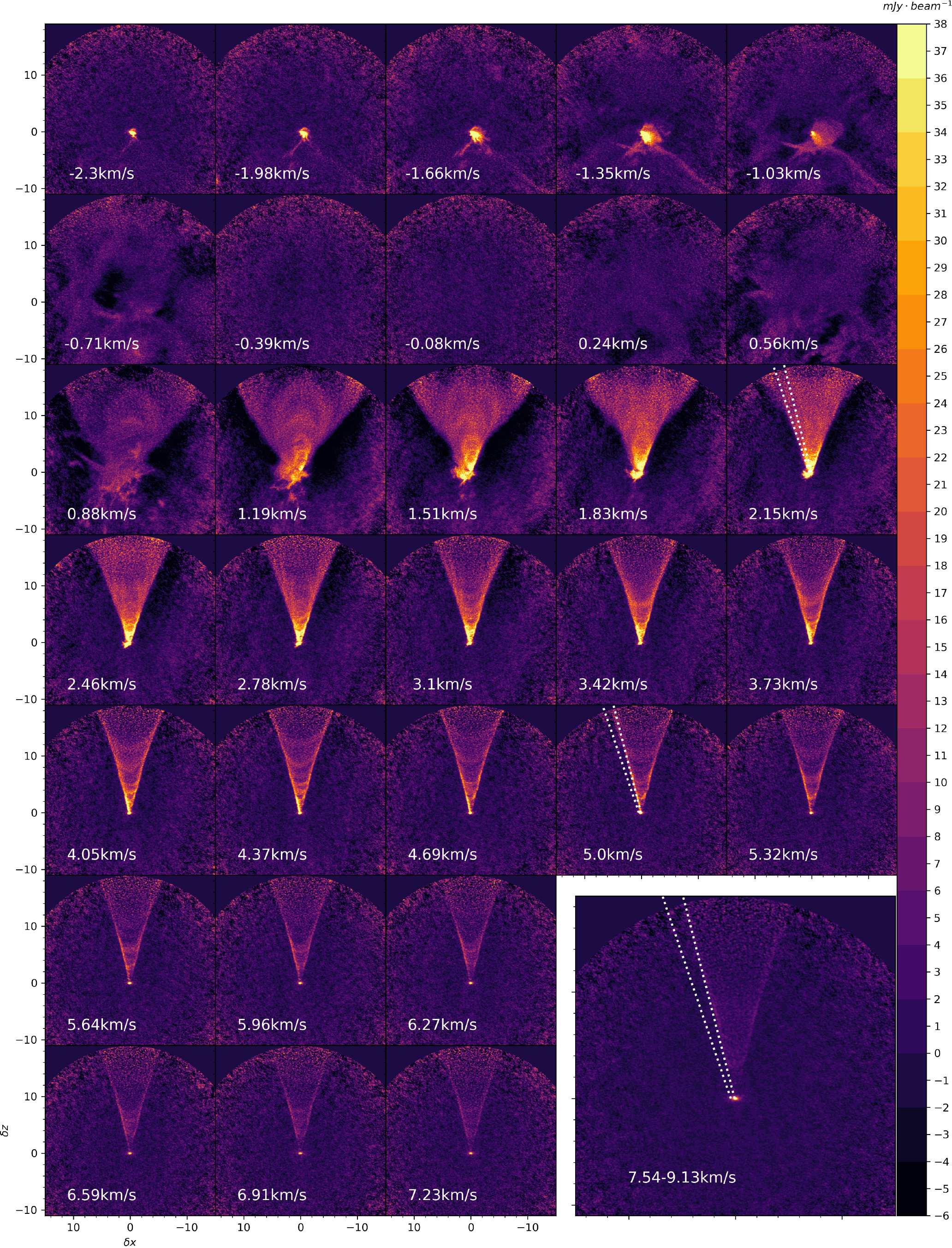}}
    \caption{\al{$^{12}$CO \COI{selected} individual channel maps. For each channel map $\Delta V = 0.32$~\kms~and the central velocity is indicated. The bottom right panel represents an averaged map from $(V-V_{\rm sys})=7.54$ to $(V-V_{\rm sys})=9.13$ \kms. The two white dotted lines show the inner and outer cone as defined in Fig.\ref{fig:XVcone}.}}
    \label{fig:CO_channels}
\end{figure*}

\section{Estimation of the mass flux}
\label{sec:Mass_flux}

\begin{figure*}
    \centering
    \resizebox{\hsize}{!}{\includegraphics{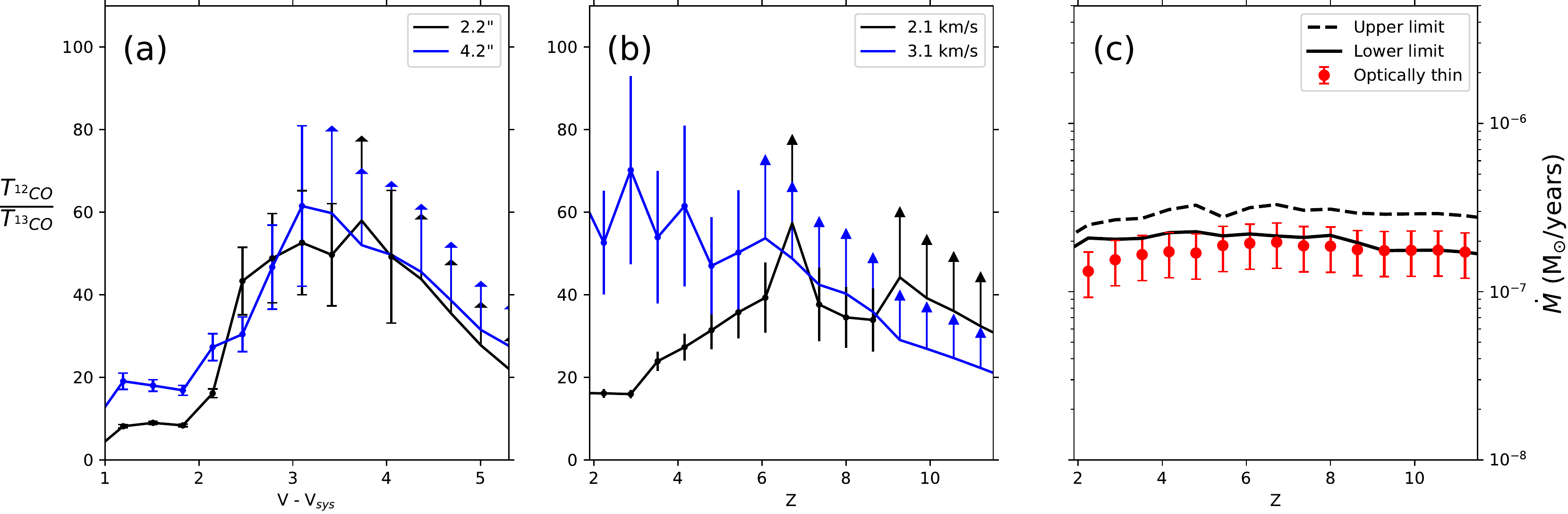}}
    \caption{\al{Intensity ratio between $^{12}$CO and $^{13}$CO \COI{radially averaged over the cone region }: a) as a function of the velocity on two slices of width = 0.7$^{\prime\prime}$ at $\delta z$ =2.2$^{\prime\prime}$~(black) and $\delta z$ = 4.2$^{\prime\prime}$~(blue); and b) as a function of height in the velocity bins $(V-V_{\rm sys}) = 2.15$ \kms~(blue) and $(V-V_{\rm sys}) = 3.1$\kms~(black). The arrows represent the 5$\sigma$ lower limit on the ratio when no $^{13}$CO is detected. c) Resulting mass flux under different opacity hypothesis assuming optically thin $^{12}$CO (in red with errorbars at 30\%), applying opacity correction based on $^{12}$CO/$^{13}$CO and \COI{assuming either optically thin emission (lower limit: black solid) or the last detected value (upper limit: black dashed) when no $^{13}$CO is detected (see text for more details).}}}
    
    \label{fig:Mdot}
\end{figure*}

\al{We calculated the mass in the conical component assuming LTE excitation and by using the method presented in \citet{louvet_hh30_2018}. We integrated $T_{\rm mb}$ from  $(V-V_{\rm sys}) = 2.15$ to 8~\kms\ in slices of $\delta z =$ 0.7$^{\prime\prime}$ \SCC{inside} the region delimited by the outer cone (see Fig. \ref{fig:XVcone}). We chose this velocity range and spatial domain to have a decent estimate of the mass without being too contaminated by the wider component at lower velocities. 
We set the CO excitation temperature to $T_{\rm ex}=50$K, corresponding to the peak $T_{\rm mb}$ in this region.  \SCC{The integrated flux over the cone area in the range $(V-V_{\rm sys}) = 2.15$ to 8~\kms\ 
and $\delta z = 2^{\prime\prime}-10^{\prime\prime}$ is 60 Jy \kms, corresponding to a total optically thin mass of $1 \times 10^{-4} M_{\odot}$\al{. A similar integrated flux was found on the SMA data of \citet{zapata_kinematics_2015}} \qu{(64 Jy \kms)}}. For a twice lower $T_{\rm ex}$, the mass would be decreased by a factor 1.6. The mass scales linearly with $T_{\rm ex}$ for $T_{\rm ex} \ge 50$K}. \al{If we integrated down to $(V-V_{\rm sys}) = 1$~\kms~, encompassing all redshifted lobe emission, the derived mass would typically increase by a factor of 1.5, but it would suffer from  contamination from the wider "pedestal" component
projected in front of the cone.}

\al{Figures \ref{fig:Mdot}a,b \SCC{show that} the intensity ratio of $^{12}$CO to $^{13}$CO \SCC{\COI{increases} as a function of} the velocity and height in the outflow.} \qu{We used the compact and more sensitive $^{13}$CO configuration for more precision on this variation.} 
\al{We added a \COI{standard} correction factor for the $^{12}$CO optical depth, which was computed for each slice \COI{in $\Delta$z }and the velocity bin. \COI{The observed ratio is always above 10 so that the  $^{13}$CO emission is optically thin. The correction factor \COI{to be applied to the $^{12}$CO brightness} is then given by :}}

\noindent
   \begin{equation}
      F_{12}=X_{12,13} \frac{T_{13CO}}{T_{12CO}},\end{equation}

\COI{with $X_{12,13,}$ the abundance ratio set at 65 \citep{goldsmith_largescale_2008}}.
\al{When no $^{13}$CO is detected, a lower and upper limit to the mass flux are obtained by assuming optically thin $^{12}$CO and by setting the \al{intensity ratio to the 
upper measured value}.}
\al{The derived \al{lower and upper limits} on the mass flux are constant from  $\delta z$=2$^{\prime\prime}$ to $\delta z$=12$^{\prime\prime}$ at \al{$\dot{M}=1.7-2.9 \times 10^{-7} M_{\odot} yr^{-1}$} (see Fig. \ref{fig:Mdot}c).}


\section{\SCC{Rotating infall} models}
\label{sec:Ulrich}

For an axisymmetric rotating envelope \SCC{in ballistic infall onto a point mass $M$}, the equation defining one given streamline is described by only two parameters: $r_d$ the outer disk radius and $\theta_0$ the initial infall angle with the rotation axis of the disk \citep{ulrich_infall_1976}. In defining $\theta$, the angle of the streamline with respect to the rotation axis at a given spherical radius $R$, we get:
\noindent
   \begin{equation}
      \cos{\theta} = \frac{-r_d \cos{\theta_0}\sin^2{\theta_0}}{R} + \cos{\theta_0} \,.\end{equation}

 We can then, with $z=r \cos{\theta}$, describe the shape of the flow surface followed by infalling material originating from $\theta_0$. The velocity components at a given point $(z,R) = (\theta,R)$ are given by:

   \begin{eqnarray}
    v_R    & = & -\bigg(\frac{GM}{R} \bigg)^{\frac{1}{2}} \bigg(1+\frac{\cos{\theta}}{\cos{\theta_0}} \bigg)^{\frac{1}{2}}\\
    v_{\theta}   & = & \bigg(\frac{GM}{R} \bigg)^{\frac{1}{2}} (\cos{\theta_0}-\cos{\theta}) \bigg(\frac{\cos{\theta_0}+\cos{\theta}}{\cos{\theta_0}\sin^2{\theta}} \bigg)^{\frac{1}{2}}\\
      v_{\phi}  & = & \bigg(\frac{GM}{R} \bigg)^{\frac{1}{2}} \frac{\sin{\theta_0}}{\sin}\bigg(1-\frac{\cos{\theta}}{\cos{\theta_0}} \bigg)^{\frac{1}{2}}.\end{eqnarray}

The paper by \citet{ulrich_infall_1976} had a \SCC{typographical error}  in the equation of $v_{\theta,}$ which was corrected by \citet{terebey_collapse_1984} (see their Eq. (89)).

We then consider at each $z(R),$ a ring of streamlines with azimuth $\phi$ from 0 to 2$\pi.$ We projected their position onto the plane of the sky and their velocity onto the line of sight, and we convolved in 3D to produce a synthetic datacube matching the angular and spectral resolutions of our ALMA observations.
We show below channel maps predicted by this model for different values of $r_d$ and $\theta_0$. 



\begin{figure*}
\centering
{
    \resizebox{\hsize}{!}{\includegraphics[width=0.9\textwidth]{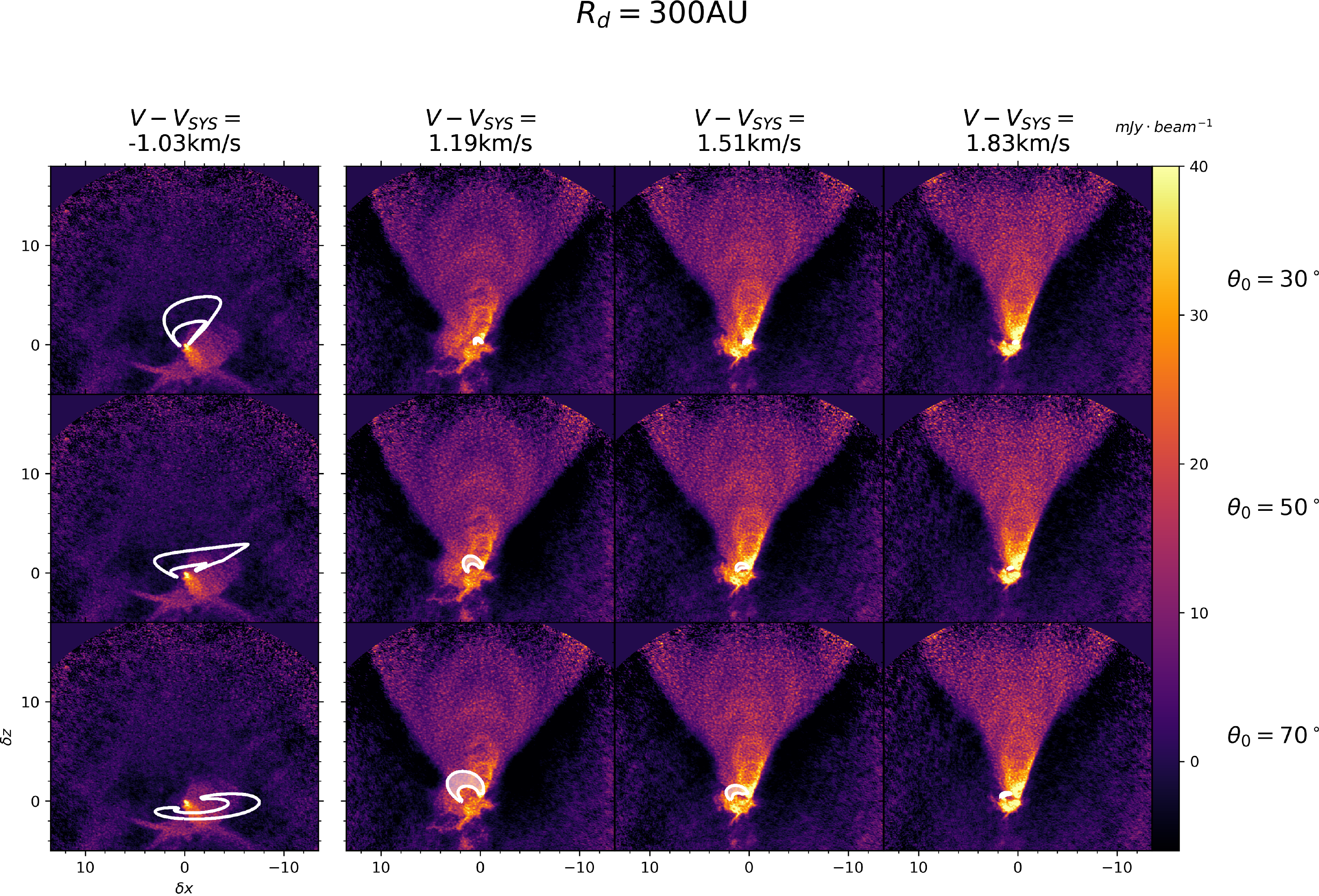}
    }
    \resizebox{\hsize}{!}{\includegraphics[width=0.9\textwidth]{InfallRd_300bis.pdf}
    }
}
\caption{$^{12}$CO \COI{selected} \COI{individual} channel maps. \COI{For each channel map $\Delta V = 0.32$~\kms~and the central velocity is indicated.} Similarly to Fig. \ref{fig:OuterOutflow}, the white contours trace the model of an infalling shell with $r_d = 300 $~au (top), 700~au (bottom), and $\theta_0=30^{\circ},50^{\circ}$, and $70^{\circ}$. }
\label{fig:Infall}

\end{figure*}



\end{appendix}
\end{document}